\documentclass[fleqn,usenatbib]{mnras}
\pdfoutput=1
\usepackage{newtxtext,newtxmath}
\usepackage[T1]{fontenc}
\usepackage{ae,aecompl}

\usepackage{graphicx}	% Including figure files
\usepackage{amsmath}	% Advanced maths commands
\usepackage{amssymb}	% Extra maths symbols

\usepackage{graphicx}	% Including figure files
\usepackage{amsmath}	% Advanced maths commands
\usepackage{amssymb}	% Extra maths symbols

\usepackage{aas_macros}
\usepackage{color}

\hypersetup{draft}
\definecolor{orange}{rgb}{1,0.5,0}
\definecolor{violet}{rgb}{0.95,0.52,0.95}

\newcommand{\rvir}{$r_\text{vir}$}
\newcommand{\rdm}{$r_\text{DM}$}
\newcommand{\rhot}{$r_\text{hot}$}
\newcommand{\rshot}{$r_\text{s,hot}$}

\newcommand{\sag}{\textsc{sag}}
\newcommand{\sagb}{\textsc{sag$_{\beta1.3}$}}

\title[Semi-Analytic Galaxies II]{Semi-Analytic Galaxies - II. Revealing the role of environmental and mass quenching in galaxy formation}

\author[Sof\'ia A. Cora et al.]{
Sof\'ia A. Cora,$^{1,2,3}$\thanks{E-mail: sacora@fcaglp.unlp.edu.ar}
Tom\'as Hough,$^{1,2,3}$
Cristian A. Vega-Mart\'inez$^{1,3}$ and  
\'Alvaro A. Orsi$^{4}$ 
\vspace{0.2cm}\\
$^{1}$Instituto de Astrof\'isica de La Plata (CCT La Plata, CONICET,  UNLP), 
   Observatorio Astron\'omico, Paseo del Bosque,\\ B1900FWA La  Plata, Argentina\\
$^{2}$Facultad de Ciencias Astron\'omicas y Geof\'{\i}sicas, 
   Universidad Nacional de La Plata, 
   Observatorio Astron\'omico, Paseo del Bosque,\\ B1900FWA La Plata, Argentina\\
$^{3}$Consejo Nacional de Investigaciones Cient\'ificas y T\'ecnicas
   (CONICET), Rivadavia 1917, Buenos Aires, Argentina\\
$^{4}$Centro de Estudios de F\'isica del Cosmos de Arag\'on, 
   Plaza de San Juan 1, Teruel, 44001, Spain\\
}

\date{Accepted XXX. Received YYY; in original form ZZZ}

\pubyear{2017}

\begin{document}
\label{firstpage}
\pagerange{\pageref{firstpage}--\pageref{lastpage}}
\maketitle

\begin{abstract}
We use the semi-analytic model of galaxy formation \sag~to study the relevance of mass and environmental quenching on  
satellite galaxies. 
We find that  
environmental processes dominate the star formation (SF) quenching of 
low-mass satellites ($M_{\star} \lesssim 10^{10.5}\, {\rm M}_{\odot}$), whereas 
high-mass galaxies typically quench as centrals. High-mass galaxies that remain 
actively forming stars
while being accreted are found to be mainly affected by mass quenching after their first infall.
For a given stellar mass, our model predicts SF quenching to be less efficient in low-mass haloes 
both before and after infall,
in contradiction with common interpretations of observational data.
Our model supports a two-stage scenario to explain the SF quenching.
Initially, the SF of satellites resembles that of centrals until the gas cooling rate is reduced to approximately half its value at infall.
Then, the SF fades through secular processes that exhaust the cold gas reservoir. This reservoir is not replenished efficiently due to the action of either ram-pressure stripping (RPS) of the hot gas in low-mass satellites, or feedback from the active galactic nucleus (AGN) in high-mass satellites. 
The delay times for the onset of SF quenching
are found to range from  $\approx 3\,{\rm Gyr}$ to $\approx 1\,{\rm Gyr}$
for low-mass ($M_{\star} \approx 10^{10}\, {\rm M}_{\odot}$) and high-mass ($M_{\star} \approx 10^{11}\, {\rm M}_{\odot}$) satellites, respectively. 
SF fades in $\approx 1.5\,{\rm Gyr}$, largely independent of stellar mass.
We find that the SF quenching of low-mass satellites 
supports the so-called delay-then-rapid quenching scenario. However,
the SF history of $z=0$ passive satellites of any stellar mass is better described by a delay-then-fade quenching scenario. 
\end{abstract}


%%%%%%%%%%%%%%%%%%%%%%%%%%%%%%%%%%%%%%%%%%%%%%%%%%

%%%%%%%%%%%%%%%%% BODY OF PAPER %%%%%%%%%%%%%%%%%%

\section{Introduction}

A great deal of work has been devoted in the last years to
determine the role of mass and environmental quenching
on the properties of central and satellite galaxies.
Mass quenching\footnote{Although the term ``secular quenching'' describes more accurately the meaning of what we refer to as ``mass quenching'', we adopt the latter following the convention in the literature.} refers to any internal process
determined by the galaxy stellar mass.
These
self-regulating processes, such as feedback from 
active galactic nuclei (AGN) 
\citep[e.g. ][]{Fabian12, Beckmann17}
and stellar feedback 
\citep[e.g. ][]{Hopkins14, Chan18}, 
can affect both central and satellite galaxies, and 
are thought to be responsible
for the dependence of galaxy properties on stellar
mass. 
Environmental quenching, on the other hand, 
corresponds to the decline of star formation (SF) activity of satellite galaxies due to their accretion into a 
massive dark matter halo. It involves environment-dependent physical processes,
like ram pressure stripping \citep[RPS, ][]{gg72}, tidal stripping 
\citep[TS, ][]{Merritt83}, 
thermal evaporation \citep{CowieSongaila77}, turbulent viscous
stripping \citep{Nulsen82} and
galaxy harassment \citep{Moore96}.

The dependence of the efficiency of mass quenching on environment and of environmental quenching on stellar mass  is somewhat controversial.
Mass and environmental quenching have been reported to be independent of
each other \citep[e.g.][]{Peng10, Quadri12, Muzzin12, Lin14, GaborDave15}, while
other studies conclude the opposite 
\citep[e.g.][]{Darvish16, Kawinwanichakij17}.
\citet{Darvish16} find 
that environmental quenching is more efficient for more massive galaxies 
($M_{\star} \gtrsim 5 \times 10^{10}\,{\rm M}_{\odot}$) at $z \lesssim 1$, 
whereas mass quenching gains relevance at $z \gtrsim 1$ and is
more efficient in higher density environments.
\citet{Kawinwanichakij17} 
extend the analysis to lower mass galaxies  
demonstrating that the presence of nearly all quiescent 
low-mass galaxies 
($\approx 3 \times 10^{9} - 10^{10}\,{\rm M}_{\odot}$) can be explained by 
environmental quenching. 
At fixed stellar mass, the star formation rate density (SFRD) 
declines with decreasing redshift more abruptly for galaxies in clusters than in the field 
\citep{Guglielmo15}, highlighting the importance of environmental quenching, especially since $z\sim 1.5$ \citep{nantais17}.
On-going programmes like GOGREEN 
(Gemini Observations of Galaxies in Rich Early Environments)
will contribute 
towards determining
the role of environment in the evolution of low-mass galaxies \citep{Balogh17}.

Galaxy interactions and mergers might explain the scenario in which environmental and mass quenching are not separable, i.e. in the regime of very dense environments and very massive galaxies \citep{Darvish16}.
Massive galaxies 
($M_{\star}\sim 10^{10} - 10^{11.5}\,M_{\odot}$) have 
typically experienced
around one 
major merger since $z = 1$ \citep{Xu2012a}.
Likewise, ellipticals in groups achieve their spheroidal morphology through
major mergers that take place at early epochs ($z \gtrsim 1$)
when they inhabit the progenitor haloes of these systems, and
their SF truncation occurs later when they fall into a galaxy group \citep{Feldmann11}.
Massive inflows
of gas in major mergers of star-forming galaxies trigger starbursts and feed the central supermassive black holes (BHs)
(\citealt[e.g.][]{Hopkins08, Khabiboulline14}); while starbursts consume most of the nuclear gas, 
feedback from supernovae (SNe) and AGN expels
the residual gas. 

Galaxy formation models have also explored 
the relative importance of different quenching mechanisms 
using both hydrodynamical simulations 
\citep[e.g. ][]{Bahe15, Taylor17, Dave17}
and semi-analytic models of galaxy formation 
\citep[SAMs; e.g.][]{delucia2012, Henriques17, Stevens17}. 
Both approaches predict that quenching begins once 
the cooling of gas becomes inefficient
\citep[e.g.][]{Schawinski14, Peng15}.
Thus, the cold gas disc is no longer replenished by cooling flows
and is consumed within a few Gyr. This results in a decline of the star formation rate  
(SFR) until the galaxy becomes passive.
Such scenario corresponds to the second stage in the so-called \textsl{delay-then-rapid} 
quenching scenario
proposed by \citet[][W13, hereafter]{Wetzel13}, in which the onset 
of quenching since the first infall 
can take $\approx 4\,{\rm Gyr}$ for 
low-mass 
galaxies 
($\approx 10^{10}\,{\rm M_{\odot}}$)
to $\approx 2\,{\rm Gyr}$ for 
high-mass 
ones ($\approx 10^{11}\,{\rm M_{\odot}}$). Once the SF quenching begins, 
this takes place in a short time-scale ($\lesssim 1\,{\rm Gyr}$). 

A crucial aspect in determining 
the gas cooling efficiency and, thus, the
quenching time-scales of satellite galaxies is the modelling of the hot gas removal in galaxies.
The updated version of the semi-analytic model 
of galaxy formation \sag~(acronym for Semi-Analytic Galaxies) presented in \citet[][hereafter Paper I]{Cora18a} incorporates a gradual removal of hot gas of satellites by the action of RPS and TS. RPS can also affect the cold gas disc. This improved treatment of 
environmental effects combined with a modified modelling of 
SN feedback has been successful in reproducing
several galaxy properties at both low and high redshifts.
In particular, the predicted fractions of quenched galaxies
as a function of stellar mass, halo mass and halo-centric distances
are in good agreement with \citet{Wetzel12}. To achieve this result, a power-law slope of the redshift 
dependence of the reheated and ejected mass is used following that from the 
zoom FIRE (Feedback
in Realistic Environments) hydrodynamical simulations
\citep{muratov15}. In this variant of the model, referred to as \sagb~in Paper I, the values of the rest of the free parameters obtained from the calibration process remain unchanged.

In this second paper of a series, we explore further the relevance
of mass quenching and environmental quenching on currently passive satellite 
galaxies, and determine the time-scales involved in the quenching process.

This paper is organized as follows. 
Section~\ref{sec:model} 
presents the galaxy formation model that combines a cosmological dark matter (DM) simulation with the \sagb~model; we describe those aspect of our SAM that are relevant to this work,
summarising the main conclusions of Paper I.
In Section~\ref{sec:fq-z0-zinfall}, we
analyse the 
fraction of
quenched galaxies 
at the time of first infall, and the relative role of environmental and mass 
quenching after their first infall.
In Section~\ref{sec:tq}, we discuss the quenching time of $z=0$ passive satellites that are star-forming at first infall.
In Section~\ref{sec:quenchscenario}, we propose a delay-then-fade quenching scenario, and
identify the physical processes associated with its two phases.
We present a summary of the main results and 
our conclusions in Section~\ref{sec:conclu}.

\section{Galaxy formation model}
\label{sec:model}

The galaxy formation model combines our semi-analytic model of galaxy formation and evolution
\sag~and the cosmological DM \textsc{MultiDark} simulation MDPL2,
which is part of the \textsc{CosmoSim} database\footnote{\url{https://www.cosmosim.org}}.
The MDPL2 simulation follows the evolution of $3840^3$ particles within a box of side-length $1\,h^{-1}\,{\rm Gpc}$,
with a mass resolution 
$m_\textrm{p} = 1.5 \times 10^{9}\, h^{-1}\, \textrm{M}_{\odot}$ per
DM~particle \citep{Klypin16}.  
It is consistent with
a flat $\Lambda$CDM model characterised by Planck cosmological parameters:
$\Omega_{\rm m}$~=~0.307, $\Omega_\Lambda$~=~0.693, 
$\Omega_{\rm B}$~=~0.048, $n_{\rm s}$~=~0.96
and $H_0$~=~100~$h^{-1}$~km~s$^{-1}$~Mpc$^{-1}$, where $h$~=~0.678 
\citep{Planck2013}.
The simulation provides the DM haloes and their corresponding merger trees required by \sag~to generate the galaxy population and track the evolution of galaxy properties. 
DM haloes have been identified with the
\textsc{Rockstar} halo finder
\citep{Behroozi_rockstar}, and merger trees were constructed with 
\textsc{ConsistentTrees} \citep{Behroozi_ctrees}.
DM haloes detected over the background density are referred to 
as \textsl{main host} haloes, whereas those lying 
within another DM haloes are subhaloes. 
\sag~assigns 
one galaxy to each new detected halo in the simulation.
Thus, each system of haloes contains a central galaxy associated to the main host halo and satellite galaxies. 
Those galaxies that are assigned to DM haloes that are no longer
identified by the halo finder after the merger of two haloes
are called orphan satellites. 

The model \sag~originates from the semi-analytic model presented
in \citet{springel2001} and was further modified as described
in \citet{cora2006},
\citet{lcp08}, \citet{tecce10}, 
\citet{orsi14}, \citet{munnozarancibia2015} and  
\citet{Gargiulo15}.
\sag~includes the effects of radiative cooling of hot gas,
star formation, feedback from SN explosions and AGN, and starbursts triggered by disc
instabilities and/or galaxy mergers, and features
a chemical enrichment model that 
tracks several chemical elements contributed by 
different sources (stellar winds and SNe 
Type Ia and II )
taking into account the lifetime of progenitors. 
This model has been improved by the implementation of a proper treatment
of environmental effects on satellite galaxies. The strangulation scheme,
in which the hot gas halo is removed instantly when the galaxy becomes a satellite, is replaced by gradual starvation produced by the combined action of RPS and TS.
Therefore, gas cooling 
takes place in both central and satellite galaxies.
The cold gas disc is also subject to 
RPS 
when it is not longer shielded by the hot gas halo;
it can also suffer the effects of TS.
Stellar mass loss as a result of TS is also taken into account.
The advantage of our implementation with respect to previous SAMs 
\citep{GonzalezPerez14, Henriques17, Stevens17} resides in the integration of the 
orbits of orphan galaxies 
according
to the potential well of the
host halo, taking into account mass loss by TS and dynamical friction effects, 
and in the use of
fitting formulae to estimate ram pressure (RP) as a function of halo mass,
halo-centric distance and redshift 
(Vega-Mart\'inez et al., in preparation).
The combination of these two improvements
provides values of RP consistent with the position and velocity
of satellite galaxies. 
The latest version of \sag~also considers a modification in the SN feedback scheme that includes an explicit redshift dependence
of the reheated and ejected mass,
and in the rate of BH growth during gas
cooling involved in the AGN feedback.
A detailed description of the updated version of \sag~is presented in Paper I. 
We summarize here the implementations related to the new SN feedback scheme, the modification of AGN feedback and the treatment of environmental effects on the hot and cold gas phases
of satellite galaxies.

\subsection{SN feedback}

SNe are associated to
star-forming
events and reheat part of the cold gas
disc. As a result, the reheated gas is transferred to the hot phase.
The reheated mass is given by
\begin{equation}
\Delta M_{\rm reheated} = \frac{4}{3} \epsilon {\frac{\eta E_\text{SN}}{V_{\rm vir}^2}} \,(1+z)^{\beta}\,\left(\frac{V_{\rm vir}}{60\,{\rm km}\,{\rm s}^{-1}}\right)^{\alpha_\text{F}}\Delta M_{\star},
\label{eq:feedfire}
\end{equation}
where $\eta$ is the number of
SNe generated from the stellar population of mass $\Delta M_\star$ formed, 
$E_\text{SN}=10^{51}\,{\rm erg}$ is the energy released
by a single SN, and $V_{\rm vir}$ is the virial velocity of the host (sub)halo.
The exponent $\alpha_\text{F}$ takes the values $-3.2$ and $-1.0$ 
for virial velocities
smaller and larger than $60\,{\rm km}\,{\rm s}^{-1}$, respectively. This additional modulation with virial velocity as well as the explicit redshift dependence are included following \citet{hirschmann16}, based on the results from
the zoom FIRE hydrodynamical simulations of \citet{muratov15}.
The
SN feedback 
efficiency, $\epsilon$, 
and the power-law slope of the redshift dependence, $
\beta$, are 
free parameters that control the amount of cold gas reheated by the
energy generated by SNe as a function of time.
The number of SNe, 
$\eta$, depends on the initial mass function (IMF);
we adopt the Chabrier IMF
\citep{Chabrier03}.

During the SN feedback process, some hot gas can also be ejected to an external reservoir. Based on the energy conservation argument
presented by \citet{guo11}, the ejected hot gas mass
is given by
\begin{equation}
\Delta M_{\rm ejected}= \frac{\Delta E_\text{SN} - 0.5\,\Delta M_{\rm reheated}\,V_{\rm vir}^2}{0.5\,V_{\rm vir}^2},
\label{eq:EnergyCons}
\end{equation}
\noindent where $\Delta E_\text{SN}$ is the energy injected by massive stars, estimated as the reheated mass but with the addition of the 
mean kinetic energy of SN ejecta per
unit mass of stars formed, $0.5\,V_\text{SN}^2$, i.e. 
\begin{equation}
\Delta E_{\rm SN} = \frac{4}{3} \epsilon_\text{ejec} {\frac{\eta E_\text{SN}}{V_{\rm vir}^2}} \,(1+z)^{\beta}\,\left(\frac{V_{\rm vir}}{60\,{\rm km}\,{\rm s}^{-1}}\right)^{\alpha_{\rm F}}\Delta M_{\star}\,0.5\,V_\text{SN}^2. 
\label{eq:ejecfire}
\end{equation}
\noindent The wind velocity is given by $V_\text{SN}=1.9\,V_\text{vir}^{1.1}$ \citep{muratov15}.
The efficiency of ejection, $\epsilon_\text{ejec}$, is another free parameter of the model. 
The ejected gas mass is assumed to
be re-incorporated back onto the (sub)halo from which it was expelled and is modelled following \citet{henriques13}. 
Thus, the reincorporated mass is given by
\begin{equation}
\Delta M_{\rm reinc}= \gamma\,\Delta M_\text{ejected}\,\frac{M_{\rm vir}}{10^{10}\,{\rm M}_{\odot}},
\label{eq:reinc}
\end{equation}
\noindent where the parameter $\gamma$ is a free parameter that regulates the efficiency of the process.

\subsection{AGN feedback}

AGN feedback reduces gas cooling in large haloes, preventing them
from forming stars at late times.
This process is triggered by
gas accretion events onto super-massive BHs. 
AGN feedback has been included in \sag~by \citet{lcp08} following \citet{croton2006},
but the modelling of the BH growth was subsequently modified
(\citealt{ruiz2015}, Paper I).

BHs grow via 
gas flows to the galactic core
triggered by disc instabilities or galaxy
mergers. BHs are assumed
to merge instantaneously when a merger occurs.
The resulting BH grows through cold gas accretion
following 
\begin{equation}
\Delta M_{\rm BH} = f_{\rm BH} \frac{{M_{\rm sat}}}{M_{\rm cen}} \frac{M_{\rm cold,sat}
+ M_{\rm cold,cen}}{(1 + 280 \, {\rm km}\,{\rm s}^{-1} / V_{\rm vir})^2},
\end{equation}
where $M_{\rm cen}$ and $M_{\rm sat}$ are the masses of the merging central and
satellite galaxies, and $M_{\rm cold, cen}$ and $M_{\rm cold,sat}$ are their
corresponding cold gas masses. 
The fraction of cold gas accreted onto the central
super-massive BH, $f_{\rm BH}$, is a free parameter of the model.

BHs can also grow during gas cooling processes taking place once a
static hot gas halo has formed around the central galaxy. Following \citet{henriques_mcmc_2015}, the BH growth rate is given by
\begin{equation}
\label{eq:BHgrowth-gascool}
\dot{M}_{\rm BH} =
 \kappa_{\rm AGN} \frac{M_{\rm BH}}{10^8 \,{\rm  M}_{\odot}}
\frac{M_{\rm hot}}{10^{11} \,{\rm  M}_{\odot}},
\end{equation}
where $M_{\rm hot}$ and
$M_{\rm BH}$ are 
the hot gas and BH masses, respectively.
The efficiency of cold gas accretion is given by the free parameter $\kappa_{\rm AGN}$.

AGN feedback reduces the amount of gas that cools by
\begin{equation}
\dot{M}_{\rm cool}^{\prime}= \dot{M}_{\rm cool}-\frac{L_{\rm BH}}{{\rm V}_{\rm vir}^2/2},
\label{eq:cool}
\end{equation}
where the BH luminosity, ${L_{\rm BH}}$, i.e. the
mechanical heating generated by the BH accretion, is given by
\begin{equation}
L_{\rm BH}= \eta \, \dot{M}_{\rm BH} \, c^2,
\label{eq:lumbh}
\end{equation}
where $\eta=0.1$ is the standard efficiency of energy
production that occurs in the vicinity of the event horizon, and $c$ is the speed of light.

\subsection{Gradual removal of the hot gas halo of satellite galaxies}

The hot gas halo of satellite galaxies is gradually stripped by the action of~RPS
and/or~TS. 
The gas beyond a
satellite-centric radius~$r_\text{sat}$ will be removed by RPS if the
pressure term $P_{\rm ram}$ satisfies the 
condition derived from the hydrodynamic simulations 
by \citet{mccarthy2008}
\begin{equation}\label{eq:rpshot}
P_{\rm ram} > \alpha_\text{RP} \frac{G
    M_\text{sat}(r_\text{sat}) \rho_\text{hot}
    (r_\text{sat})}{r_\text{sat}}.
\end{equation}
\noindent Here, $\rho_\text{hot}$ is the density profile of the hot gas halo for an
isothermal sphere, i.e.,  
$\rho_\text{hot} = {M_\text{hot}}/({4\pi\, r_\text{hot}\, r^2})$, 
where $r_\text{hot}$ is
the radius containing $M_\text{hot}$.
This radius 
initially adopts the value of the subhalo 
virial radius, \rvir. In the case of orphan satellites, \rvir~preserves
the value corresponding to the last time the subhalo was identified.
For the geometrical constant $\alpha_\text{RP}$ we adopt $\alpha_\text{RP}=5$, which is an intermediate value within the range considered by \citet{mccarthy2008}.
The total mass $M_\text{sat}$ of a satellite is
\begin{equation}\label{eq:msat}
  \begin{split}
    M_\text{sat}(r_\text{sat}) = & M_\star + M_\text{cold}\\
    &+ 4\pi \int_0^{r_\text{sat}} \left[ \rho_\text{hot}(r) +
    \rho_\text{DM}(r) \right] r^2 dr,
  \end{split}
\end{equation}
assuming that $r_\text{sat}$ is large enough to contain all the stars and cold
gas. 
We also assume that the DM 
is distributed in
an isothermal sphere density profile.
The hot gas stripping radius due to~RP, 
$r_\text{s,hot}^\text{RPS}$ can be obtained by solving numerically the combined equations ~\eqref{eq:rpshot} and~\eqref{eq:msat} (see Appendix A in Paper I).
This radius is compared with the
tidal radius determined by TS,
that is,
$r_\text{s,hot}^\text{TS}=r_\text{DM}$,
where the bounding radius for the DM,~\rdm, 
is given by \rvir; we are assuming that 
the hot gas distributes
parallel to the~DM.
The smaller of 
$r_\text{s,hot}^\text{RPS}$ and $r_\text{s,hot}^\text{TS}$
is the stripping radius~\rshot.
Thus, the value of~\rhot~is updated such that 
$r_\text{hot}^\text{new}=r_\text{s,hot}$.
All hot gas beyond that radius is stripped.
The
remaining gas is redistributed restoring an isothermal profile,
but truncated at~\rhot,
as in~\citet{font2008} and \citet{kimm2011}.

\subsection[]{Ram pressure and tidal stripping of cold gas disc}
\label{sec_RPS_TS_cold}

RPS of cold gas is modelled according to the simple criterion proposed by \citet{gg72}, which was implemented in \sag~by \citet{tecce10}. The cold gas of the
galactic disc located at a galactocentric radius~$R$ is stripped away when
the~RP exerted on the galaxy by the intragroup/intracluster medium exceeds the restoring force
per unit area due to the gravity of the disc, that is
\begin{equation}\label{eq:rps}
  P_{\rm ram} > 2\pi G \,\Sigma_\text{disc}(R)\,
  \Sigma_\text{cold}(R), 
\end{equation}
where $\Sigma_\text{disc}$, $\Sigma_\text{cold}$ are the surface densities of the
galactic disc (stars plus cold gas) and of the cold gas disc, respectively.
The discs of stars and gas are modelled by 
an exponential surface density profile given by
$\Sigma(R) = \Sigma_0 \exp(-R/R_\text{d})$
where $\Sigma_0$ is the central surface density and~$R_\text{d}$ is the  scale length
of the
disc. This scale length is estimated as  
$R_\text{d}=(\lambda/\sqrt{2})R_\text{vir}$ \citep{mmw98},
where $\lambda$ is the spin parameter of the DM halo in which
the galaxy resides.
The stripping radius~$R_\text{s,cold}^\text{RPS}$, beyond which all cold gas is removed, is derived from Eq.~\eqref{eq:rps} and is given by 
\begin{equation}\label{eq:rstrip}
  R_\text{s,cold}^\text{RPS} = -0.5 R_\text{d} \ln \left( \frac{P_{\rm ram}}{2\pi G \,\Sigma_\text{0,disc}\,
  \Sigma_\text{0,cold}} \right),
\end{equation}
where $\Sigma_\text{0,disc}$ and $\Sigma_\text{0,cold}$ are the central surface
densities of the stellar disc and of the cold gas disc, respectively. For simplicity, both disc components are considered to have the same scale length.
We have checked that the precise distribution of the stellar content has a  negligible impact on our results.

We assume that the hot gas halo of satellite galaxies shields the cold gas
disc from the action of~RPS. Thus, the ambient~RP starts affecting the cold gas once 
the hot gas halo has been 
reduced to very low values by gas cooling and/or stripping processes (RPS, TS), that is, when its mass becomes less than $10$ percent the baryonic mass of the galaxy.
This threshold has been chosen small enough to allow the role of the hot gas as a 
shield for a sufficiently long time, otherwise RPS on the cold disc gas
becomes too effective.

The stripping radius due to TS, 
$R_\text{s,cold}^\text{TS}$, is determined by the bounding radius of the DM halo, ~\rdm.
At each snapshot of the simulation, $R_\text{s,cold}^\text{TS}$ is compared with  
the stripping radius determined by RPS, $R_\text{s,cold}^\text{RPS}$ (equation~\eqref{eq:rstrip}), in order 
to account for the effect of tides. 
The smaller of these two radii determines the stripping radius $R_\text{s,cold}$ and all the gas beyond this radius
is stripped away.
The stripped gas is added 
either to the hot gas component of 
the central galaxy (which represents the intragroup/intracluster medium) or to the hot gas halo of satellite galaxies  
(an orphan galaxy could in turn orbit 
a satellite galaxy). 
If the hot gas phase is no longer present in the latter case, the stripped cold gas is transferred
to the hot gas contained in the main halo.

After a stripping event produced by RPS and/or TS, the
remaining disc gas is redistributed following an exponential surface 
density profile truncated at $R_\text{cold}^\text{new}=R_\text{s,cold}$,
 with a new scale length defined as 
$R_\text{d,cold}^\text{new}=R_\text{cold}^\text{new}/7$ 
(assuming that $99$ percent of the cold gas disc is contained within $7*R_\text{d,cold}$).

\subsection{Galaxy properties from the SAG$_{\beta 1.3}$ model}

The free parameters of 
the \sag~model
were tuned using the Particle Swarm 
Optimization Technique (PSO) by \citet{ruiz2015} adopting a set of five observational constraints defined in 
\citet{Knebe18}. Namely, these are the stellar mass
functions (SMF) at $z=0$ and $z=2$ 
(compilation data used by \citealt{henriques_mcmc_2015}), the star formation rate (SFR) distribution
function for the
redshift interval $z \in [0.0,0.3]$ \citep{gruppioni15}, 
the fraction of mass in cold gas as a function of stellar mass 
\citep{boselli14}, 
and the relation between bulge mass and the mass of the central
supermassive black hole, taken from \citet{mcconnell_bhb_2013} and \citet{kormendy_bhb_2013}. 
During the calibration process, the power-law slope $\beta$ that
characterises the redshift dependence included in the modified SN feedback scheme 
(eqs.~\ref{eq:feedfire} and \ref{eq:ejecfire})
was considered a free parameter of the model.
As discussed in Paper I, the fit to the SMF at $z=2$ results in $\beta=1.99$, which is above the value obtained from the zoom FIRE hydrodynamical simulations of 
\citet{muratov15}, $\beta=1.3$.
As a result, the predicted SFRD
becomes too low with respect to observational data
at high redshifts. A better agreement between model predictions and
observational data is obtained by fixing $\beta=1.3$ while keeping
the rest of the parameter values from the calibration process.
This variant of the model is referred to as the \sagb~model in Paper I. As a result of this choice for $\beta$, the model predicts a higher number density of low-mass galaxies in the SMF at $z=2$.

We classify galaxies as 
star-forming
or passive according to their
specific star formation rate density (sSFR).
Passive galaxies are those with 
${\rm sSFR} < 10^{-10.7}\,{\rm yr}^{-1}$. This cut is obtained
from the distribution of sSFR of our model galaxies 
(see fig.~9 in Paper I).
The underpredicted SFRD at high redshifts from the calibrated model (characterised by $\beta=1.99$) results in
fewer
passive galaxies at $z=0$ (see fig.~10 in Paper I). This occurs due to
the bulk of SF being shifted to lower redshifts, meaning that
the galaxies lack sufficient time to quench their SF.
Only the most massive galaxies
($\log(M_{\star}[{\rm M}_{\odot}])\in [10.9, 11.3]$),
within DM host haloes with masses
$M_{\rm halo}\gtrsim 10^{14}\,{\rm M}_{\odot}$,
achieve quenched fractions similar to those presented by \citet{Wetzel12}.
By using a milder redshift dependence of the reheated and ejected mass, as characterised by \sagb, we achieve the
expected behaviour of the fractions of quenched galaxies
as a function of stellar mass, halo mass and the halo-centric distances
(see figs.~11 and 12 in Paper I).
The analysis of the galaxy catalogue generated by the \sagb~model
allows us to conclude that RPS plays a dominant role
among the environmental processes considered in our model, being more effective for lower mass galaxies 
residing in more massive haloes
and at lower redshifts
(see figs.~13, 14 and 15 in Paper I), and 
adequately regulates
the mass of the hot gas halo and the cold gas disc.
While the gradual starvation of the hot
gas reservoir is a key ingredient
to determine the right SF history of satellite galaxies, 
the mild effect suffered by the cold gas disc improves the atomic gas content of galaxies when compared to observations (see fig.~16 in Paper I).
In view of these results, the current work is based on the analysis of
the properties of galaxies generated by the \sagb~model.

\section{Quenched fractions: 
dependence on
time at first infall}
\label{sec:fq-z0-zinfall}

Both mass and environmental quenching can affect the galaxy
after being accreted but their relative role depends
on both stellar mass and host halo mass
and cannot be estimated from the fraction of
quenched galaxies at $z=0$.
We 
explore this issue 
by
making a deeper analysis of further information
provided by the \sagb~model. This involves the 
dependence 
of quenched fractions
and stellar mass growth 
on
the time of first infall,
defined as the moment in which the galaxy
becomes a satellite for the first time; there is strong
evidence that environmental effects
produce SF quenching 
from
first infall (W13). 
We denote the corresponding redshift as $z_{\rm infall}$. \\

\subsection{$z=0$ quenched fractions}
\label{sec:fqz0}

\begin{figure}
  \centering
  \includegraphics[width=\columnwidth]{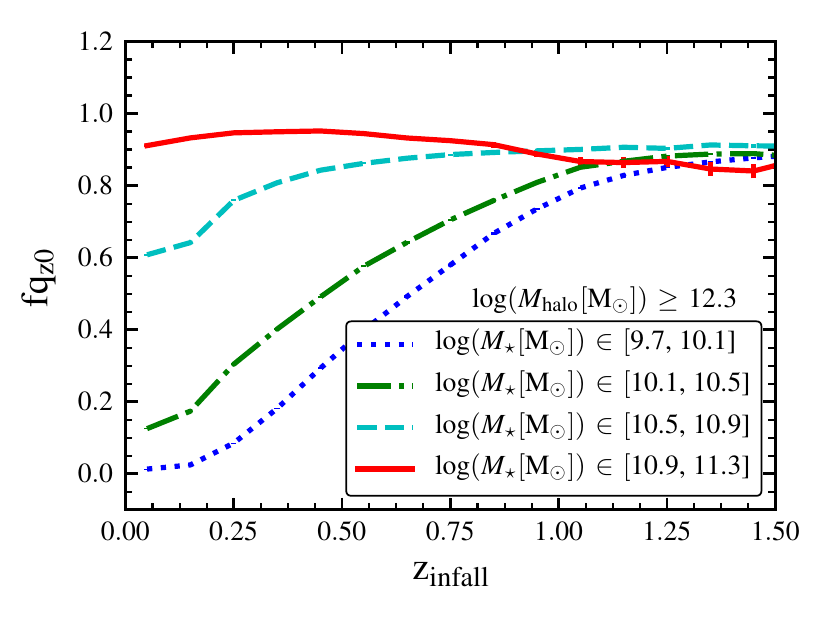}
  \caption{
Fraction of
satellite galaxies that are quenched at $z=0$, $fq_{\rm z0}$,
as a function of the redshift at infall, $z_{\rm infall}$.
All galaxies identified as satellites within main host haloes of present-day mass
${\rm log} (M_{\rm halo} [{\rm M}_{\odot}]) \geq 12.3$
are included.
Different lines represent the values of $fq_{\rm z0}$ 
for galaxies in different
local stellar mass ranges, 
as indicated in the legend. 
Error bars show the $68$ percent bayesian confidence interval estimated
following \citet{Cameron11};
they are only visible for the highest stellar mass bin. 
}
  \label{fig:fqz0_zinfall}
\end{figure}

Fig.~\ref{fig:fqz0_zinfall} shows the fraction of 
quenched satellites at $z=0$, $fq_{\rm z0}$, 
as a function of $z_{\rm infall}$.
Satellites are binned
according to their present-day stellar mass as indicated in the legend. 
We consider those galaxies identified as satellites within main host haloes of mass 
$M_{\rm halo} \geq 10^{12.3}\,{\rm M}_{\odot}$, 
following the selection criterion adopted in \citet{Wetzel12}. 
For a given stellar mass range,
$fq_{\rm z0}$ increases for satellites with higher values of $z_{\rm infall}$, as expected. 
Thus, galaxies that have been satellites for longer have had more time
to experience 
environmental quenching. 
Values of $fq_{\rm z0}$ are larger 
for larger stellar masses, at any $z_{\rm infall}$.
However, the dependence of $fq_{\rm z0}$ 
on
$z_{\rm infall}$
is more pronounced for less
massive satellites. This trend becomes gradually milder for more
massive galaxies.
For a given stellar mass range, 
the average values of the quenched fractions are 
consistent with the results 
of the stellar and halo mass dependence of $fq_{\rm z0}$
shown in 
fig.~11
of Paper I.
The relations $fq_{\rm z0}$ vs. $z_{\rm infall}$ are separated by a major gap
between the two lowest and two highest stellar mass bins
which is produced by the under-prediction of passive galaxies in the mass
range ${\log} (M_{\star} [{\rm M}_{\odot}]) \in [10.1, 10.5]$, 
discussed in Paper I. 

\begin{figure*}
  \centering
  \includegraphics[width=2.0\columnwidth]{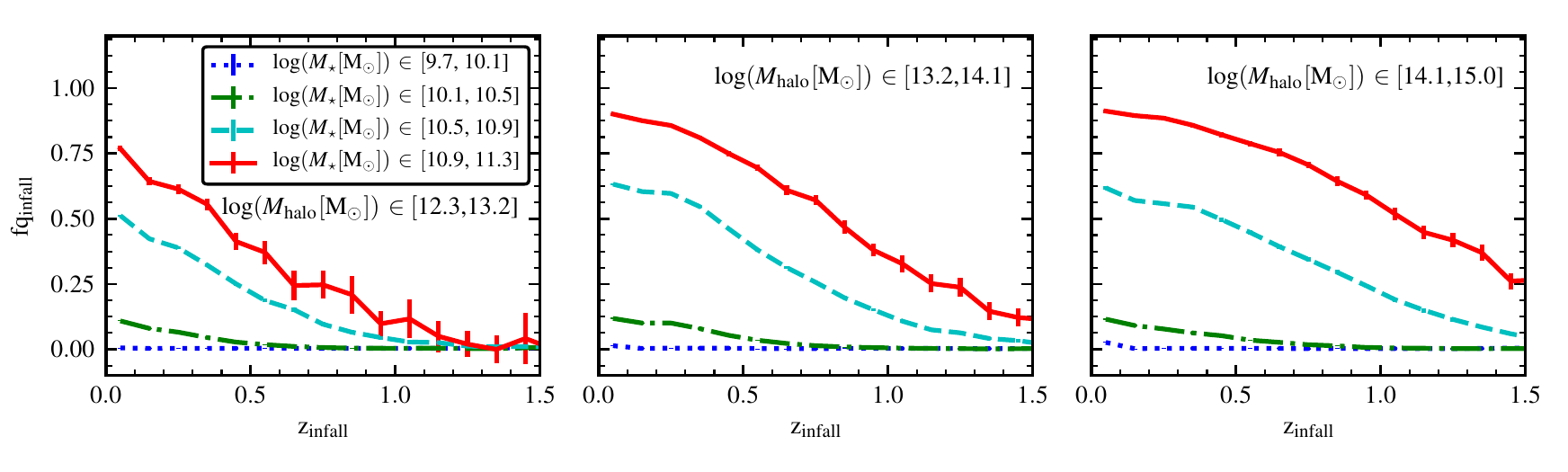}
  \caption{
Fraction of
satellite galaxies that are quenched at first infall, $fq_{\rm infall}$,
as a function of the redshift at infall, $z_{\rm infall}$.
Different lines represent the values of $fq_{\rm infall}$
for satellites in different
present-day stellar mass ranges, as indicated in the legend.
They are grouped according to the mass of their main host haloes 
(different panels). 
Error bars show the $68$ percent bayesian confidence interval estimated
following \citet{Cameron11};
they are only visible for the highest stellar mass bin. 
}
  \label{fig:fqzinfall_zinfall}
\end{figure*}

Low-mass
satellites characterised by  
$z_{\rm infall} \approx 0$ are mostly 
star-forming
galaxies 
($fq_{\rm z0} \approx 0.0 - 0.1$). This is because SF quenching as a result of
environmental effects is negligible for galaxies that have been recently 
accreted, and these galaxies are too small to suffer the consequences
of mass quenching 
while they were centrals.
However, 
more than $50$ percent of 
high-mass 
satellites 
accreted at the present epoch are passive.
Massive galaxies
have thus
suffered mass quenching processes as centrals prior to infall 
(e.g. AGN feedback
that reduces the gas cooling rates;
starbursts
triggered by both mergers and disk instabilities
that exhaust the cold gas reservoir)
as already found by 
\citet{vandenBosch08}. 
Hence, environmental 
processes can be considered
as the dominant quenching mechanism in low-mass 
galaxies consistent with the conclusions of \citet{Wetzel12}.

The fraction of quenched galaxies at the present epoch reach 
similarly high
values
($\sim 0.8 - 0.95$)
for all galaxies regardless of their stellar mass when they have been
satellites for more than $\approx 8\, {\rm Gyr}$ ($z_{\rm infall} \gtrsim 1$).
The time elapsed
is enough for the maximum possible
effect of those processes affecting satellites to take place. 
Both the nature of these processes
and their relative role (if there is more than one at play, e.g. environmental
quenching and/or mass quenching) 
can vary for galaxies with different masses. 
\citet{Lin14} investigate the relation
between environmental quenching and mass quenching efficiencies
with stellar mass for galaxies in
a large optically selected sample of field and group
galaxies obtained from the Pan-STARRS1 Medium-Deep
Survey, which span a redshift range $0.2 <z<0.8$.
For lower redshifts ($0.2<z<0.5$), they find that SF in more massive galaxies
($M_{\star} \gtrsim 1-2 \times 10^{10}\,M_{\odot}$)
is mainly suppressed by mass quenching processes
while environmental quenching gains relevance for lower mass galaxies. 
Mass quenching keeps the dominant role for massive galaxies
in the higher redshifts probed ($0.5 < z < 0.8$).
This range overlaps with the one considered by
\citet{Kawinwanichakij17},
who find similar results from
the FourStar Galaxy Evolution (ZFOURGE) survey.

\subsection{Quenched fractions at first infall}
\label{sec:fqinfall}

We study the fraction of satellite galaxies selected at $z=0$ that are quenched at the time of first infall, $fq_{\rm infall}$.
This is 
estimated from the SFR and stellar mass at $z_{\rm infall}$.
The relation between $fq_{\rm infall}$ and $z_{\rm infall}$
is presented in
Fig.~\ref{fig:fqzinfall_zinfall},   
grouping satellite galaxies according to their local stellar mass (different lines)
and the mass 
of their main host haloes 
(different panels). 
Low-mass 
satellites 
have not suffered mass quenching before becoming satellites,
since $fq_{\rm infall} \approx 0$ for any value of $z_{\rm infall}$.  This is consistent with the previous analysis of $fq_{\rm z0}$ (Fig.~\ref{fig:fqz0_zinfall}). Intermediate mass satellites
($M_{\star} [{\rm M}_{\odot}] \in [10^{10.5}, 10^{10.9}]$)
accreted at $z_{\rm infall} \gtrsim 1.5$
also have $fq_{\rm infall} \approx 0$, 
being almost all still 
star-forming
by that time. The most massive satellites considered in this analysis
($M_{\star} [{\rm M}_{\odot}] \in [10^{10.9}, 10^{11.3}]$) 
have $fq_{\rm infall}\approx 0 - 0.3$
for these early accretion times, with the larger fractions corresponding to those satellites 
residing within massive main host haloes 
($M_{\rm vir} [{\rm M}_{\odot}] \in [10^{14.1}, 10^{15.}]$).
As $z_{\rm infall}$ decreases, $fq_{\rm infall}$ increases
monotonically 
for galaxies in the two more massive stellar-mass bins,
These galaxies experience mass quenching processes as centrals prior to infall, which becomes more efficient due to stellar mass growth.

Combining the results of the fractions $fq_{\rm z0}$ and $fq_{\rm infall}$
for a given stellar mass, we see that for 
$M_{\star} \gtrsim 10^{10.9} \,{\rm M}_{\odot}$
almost
all satellite galaxies that are 
star-forming
at time of first infall are quenched by $z=0$,
regardless of $z_{\rm infall}$.
The fraction of 
low-mass 
satellites 
($M_{\star} \lesssim 10^{10.1} \,{\rm M}_{\odot}$)
that were 
star-forming
at infall and 
quenched after infall is directly given by $fq_{\rm z0}$
because galaxies in this mass range are all 
star-forming
at infall
($fq_{\rm zinfall}\approx 0$)
regardless of $z_{\rm infall}$. 
If satellites are not discriminated according
to $z_{\rm infall}$ then we find that $fq_{\rm z0}\approx 0.46$ for low-mass satellites residing
within main host haloes of mass 
$M_{\rm halo} \geq 10^{12.3}\,{\rm M}_{\odot}$.  
Thus, we arise to 
conclusions similar to those obtained
by W13 from the analysis of the results presented in their fig. 7.
That is, 
half of 
the
satellites with low stellar mass 
that were 
star-forming
at the time of first infall have been
quenched by the present, while
essentially all massive satellites 
that were 
initially
star-forming
have been quenched. 

The dependence of the fractions 
$fq_{\rm infall}$
on $z_{\rm infall}$ for different stellar mass ranges
only changes slightly for galaxies within haloes
of mass 
$M_{\rm halo} \gtrsim 10^{13.2}\,{\rm M}_{\odot}$,
as can be seen from comparing the middle and right panels of 
Fig.~\ref{fig:fqzinfall_zinfall}.
This lack of strong dependence of the quenched fractions 
on main host halo mass
for high-mass
haloes
is consistent with the behaviour that emerged from the analysis
of the radial dependence of the quenched fraction presented in Paper I (see fig. 
12).
The major difference 
appears
for galaxies
in the two highest stellar mass bins 
that have been accreted by 
smaller mass haloes, where
$fq_{\rm infall}$ decreases by $\sim 0.15$, 
regardless of $z_{\rm infall}$,
with respect to those 
fractions 
typical of more massive haloes. 
This is the result
of selecting galaxies by the same present-day stellar mass combined
with the different stellar-mass growth rate   
experienced by 
galaxies 
of a given $z=0$ stellar mass 
that are
located in different environments \citep{Guglielmo15};
cluster galaxies form their stars sooner than those in the field, 
giving rise to a more pronounced slope of the decline of the cosmic 
SFR
density with redshift in clusters. 
Thus,
galaxies selected within a given stellar-mass range at $z=0$ that have been accreted by lower mass haloes have smaller
stellar mass at a given time of infall 
\citep[see fig. 12 in][]{behroozi13c}, 
which implies
lower efficiency of mass quenching prior to infall with the
consequent lower values of $fq_{\rm infall}$. 

\subsection{Implications of the fraction of currently passive satellites already quenched at $z_{\rm infall}$}
\label{sec:implications}

\begin{figure}
  \centering
  \includegraphics[width=\columnwidth]{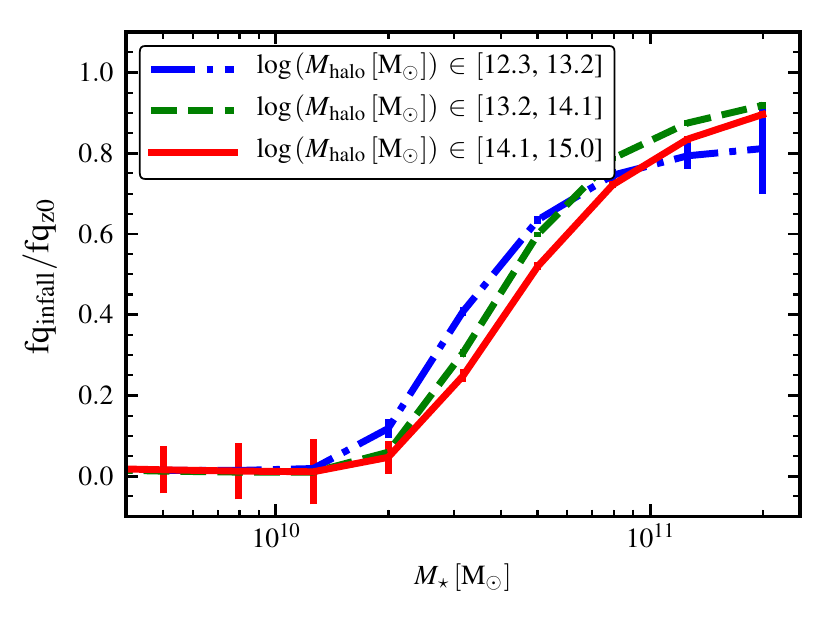}
  \caption{Fraction of 
  $z=0$ quenched satellite galaxies that are
quenched at first infall as a function of present-day stellar mass for galaxies
residing within main host haloes of different mass (different line styles). 
Error bars are obtained from the propagation of errors of the quotient.
The fractions $fq_{\rm infall}/fq_{\rm z0}$ increase with stellar mass, meaning
that more massive galaxies are more likely to be quenched
as centrals through mass quenching processes.
For intermediate mass satellites,
the ratio $fq_{\rm infall}/fq_{\rm z0}$
is larger for lower mass haloes.
}
  \label{fig:fq-at-infall-z0-ms-mh}
\end{figure}

The fraction of currently quiescent satellites
that are already quenched at first infall
can be estimated
by combining
the information provided by the 
$z=0$
quenched fraction
($fq_{\rm z0}$, Fig.~\ref{fig:fqz0_zinfall}) and the quenched fraction
at first infall ($fq_{\rm infall}$, Fig.~\ref{fig:fqzinfall_zinfall}).
Thus, this fraction is calculated simply 
by the ratio between
$fq_{\rm zinfall}$
and $fq_{\rm z0}$.
Fig.~\ref{fig:fq-at-infall-z0-ms-mh} shows the ratio 
$fq_{\rm infall}/fq_{\rm z0}$ as a function of stellar mass in bins of 
present-day main host halo mass.
This ratio increases with stellar mass, meaning
that more massive galaxies are more likely to be quenched
as centrals through mass quenching processes, consistent with 
the conclusion inferred from 
Fig.~\ref{fig:fqzinfall_zinfall}. 
At fixed stellar mass (within the range 
$M_{\star} [{\rm M}_{\odot}] \in [10^{10.1}, 10^{10.5}]$),
this ratio is larger for satellites residing in lower mass haloes.

The dependence of
the ratio $fq_{\rm infall}/fq_{\rm z0}$
on stellar mass and halo
mass is shown in Fig.~\ref{fig:fq-at-infall-z0-ms-mh}. Our model predictions are 
similar to W13 
(see their fig. 10a), but the halo-mass dependence is less pronounced for \sag~model galaxies.
Their results are obtained from 
the comparison of 
galaxy group/cluster catalogues
from the Sloan Digital Sky Survey (SDSS) Data Release 7 with
a galaxy catalogue
constructed from a high-resolution
cosmological simulation
by applying the subhalo abundance matching (SHAM) technique
\citep{Conroy06, Vale06}.
From 
the trend with halo mass,
W13
conclude that quenching prior to first infall is more important in
galaxies accreted by
lower mass host haloes,
finding this fact reasonable considering that 
they
fell in more recently. 
However,
such a trend can be interpreted in a different way if
we take into account that,
for quenched satellites 
at a fixed present-day stellar mass and epoch,
$fq_{\rm infall}$ is lower in less massive haloes because of the lower stellar mass at the time of first infall.
As a consequence of this, mass quenching is less effective prior to infall for galaxies accreted by less massive haloes, as we discuss in Section~\ref{sec:fqinfall}.
The ratio $fq_{\rm infall}/fq_{\rm z0}$ becomes larger 
for galaxies residing in lower mass haloes because 
$fq_{\rm z0}$ is also smaller for them (although not shown in Fig.~\ref{fig:fqz0_zinfall}). 
This is a result of the milder effect of RPS (found to be the dominant environmental process among those considered in \sag) exerted by 
lower mass haloes 
\citep[also see figs.~14 and 15 of][]{tecce10}. 

Hence, from the analysis of model results, 
we can assert that only considering the fraction
$fq_{\rm infall}/fq_{\rm z0}$ in the way presented in 
Fig.~\ref{fig:fq-at-infall-z0-ms-mh} 
is misleading: 
there are many variables playing an important role in the
process of SF quenching.
With the additional information provided by 
the fractions $fq_{\rm z0}$ and $fq_{\rm infall}$, separately,
we conclude that 
mass quenching prior to infall is less efficient for galaxies accreted by 
less massive haloes, and quenching after infall is also less efficient in smaller
haloes.

\subsection{Relative role of environmental and mass quenching after first infall}
\label{sec:mass_quench_after}

\begin{figure}
  \centering
  \includegraphics[width=1.0\columnwidth]{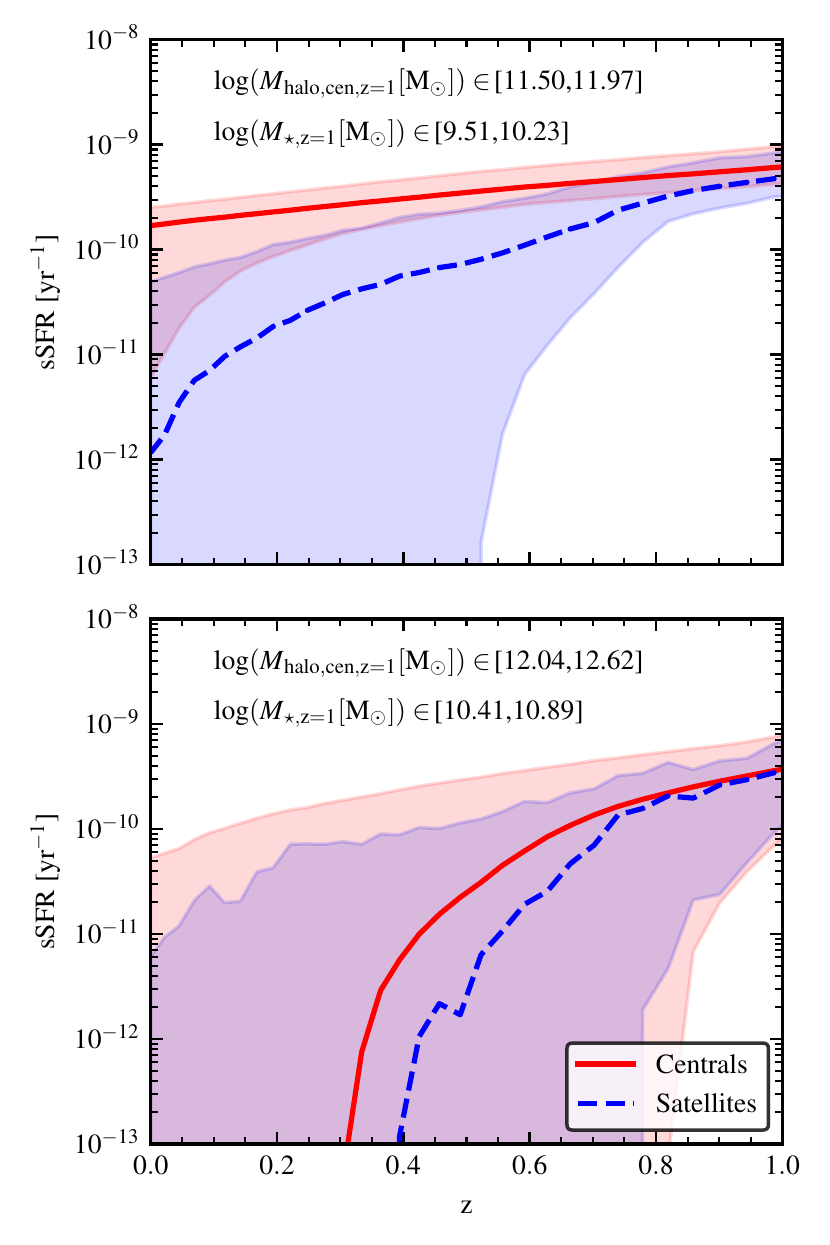}
  \caption{
Comparison of the evolution of the sSFR from
$z=1$ of current satellite galaxies that have been accreted at $z_{\rm infall}=1$ and are 
star-forming
at that time (blue dashed line) with that of a control sample built with current central galaxies that are also 
star-forming
at $z=1$ (red solid line).
Lines represent median values and the corresponding 
shaded
areas depict the $10$th and $90$th percentiles. 
Satellites residing in main host haloes with 
present-day mass $M_{\rm halo} \geq 12.3 \,{\rm M}_{\odot}$ are considered. They are selected according to their $z=0$ stellar mass and are tracked back till $z=1$ to define the ranges of stellar mass and subhalo mass
that characterise them at the time of first infall.
By 
calculating
the
$10$th and $90$th
percentiles of the mass distributions, we define the $z=1$ stellar mass and halo mass bins
that are considered to select the population of central galaxies, as indicated in the legend.
This selection guarantees that satellites and centrals have similar initial stellar mass and (sub)halo mass at $z=1$.
{\em Top panel:} Galaxy populations determined by
satellites with present-day stellar mass within the two less massive ranges considered in Fig.~\ref{fig:fqz0_zinfall} and
Fig.~\ref{fig:fqzinfall_zinfall}, i.e. $M_{\star} [{\rm M}_{\odot}]) \in [10^{9.7}, 10^{10.5}]$.
{\em Bottom panel:}
Galaxy populations determined by satellites with 
$z=0$
stellar mass within the two more massive ranges considered, i.e. 
$M_{\star} [{\rm M}_{\odot}]) \in [10^{10.5}, 10^{11.3}]$.
}
  \label{fig:sSFRevolSatCen}
\end{figure}

\begin{figure}
  \centering
  \includegraphics[width=1.0\columnwidth]{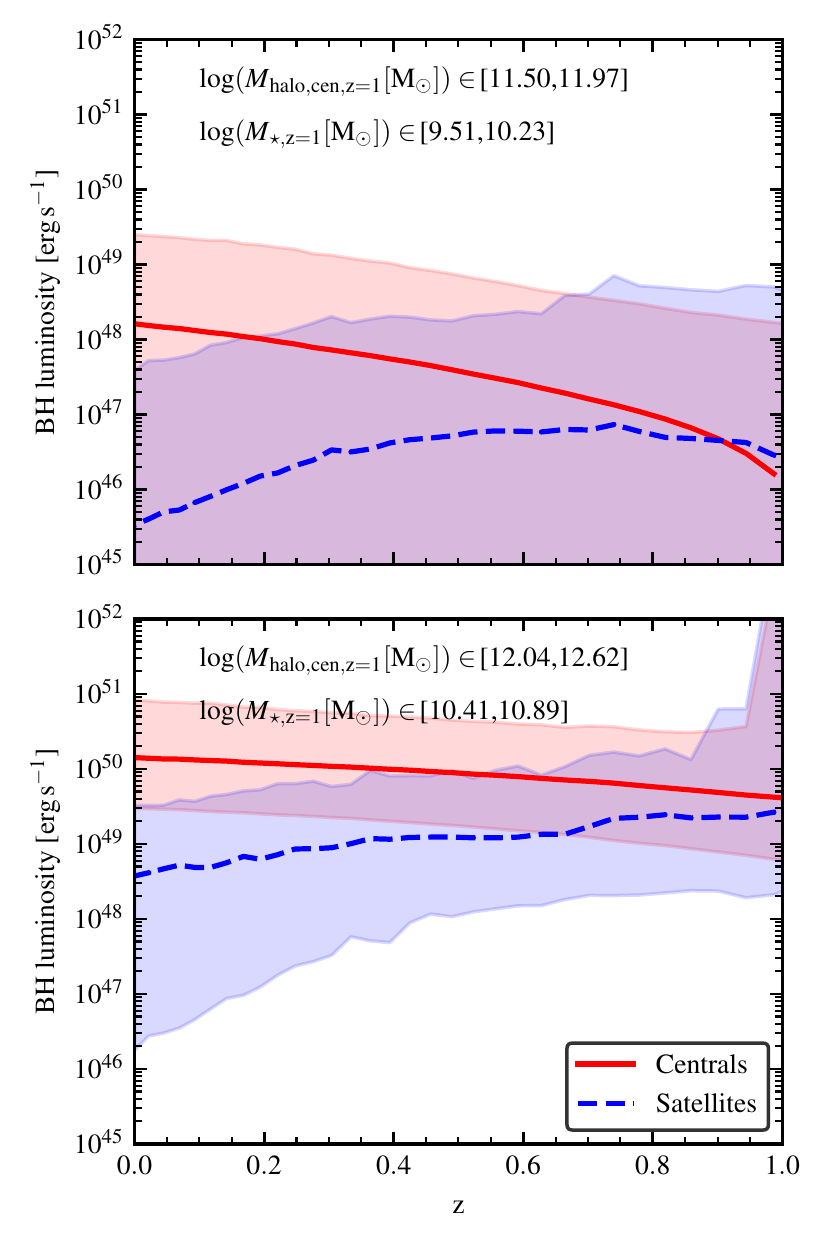}
    \caption{
Same as Fig.~\ref{fig:sSFRevolSatCen} but for the evolution of the BH luminosity.
}
  \label{fig:lumBHevolSatCen}
\end{figure}

In order to disentangle the role of environmental 
and mass quenching
after first infall for satellites
with different present-day stellar mass, 
we compare
their evolution to a control sample of central galaxies that never become satellites.
We consider satellites residing in main host haloes with 
present-day mass $M_{\rm halo} \geq 12.3 \,{\rm M}_{\odot}$.
They are selected according to their $z=0$ stellar mass, keeping those that
have been accreted at $z_{\rm infall}=1$ and were 
star-forming
at that time. 
The control sample of 
star-forming
central galaxies at $z=1$ is built with the requirements that they
have the same stellar mass as satellites at infall and reside within main host haloes with masses 
in the range of subhalo masses that characterise the satellite galaxies at the
moment of accretion.
These stellar and subhalo mass bins
are defined by 
tracking satellites back in time
till $z=1$ and 
calculating the 
$10$th and $90$th
percentiles of these mass distributions. 
This information is detailed in the legends of 
Figs.~\ref{fig:sSFRevolSatCen} and~\ref{fig:lumBHevolSatCen} that show the evolution of the sSFR and the BH luminosity (given by eq.~\ref{eq:lumbh}), respectively, for
both galaxy populations. 
Lines represent median values and the corresponding dashed areas depict the $10$th and $90$th percentiles. 
The upper panel corresponds to galaxy populations determined by
satellites with present-day stellar mass within the two less massive ranges considered in Fig.~\ref{fig:fqz0_zinfall} and
Fig.~\ref{fig:fqzinfall_zinfall}, i.e. $M_{\star} [{\rm M}_{\odot}] \in [10^{9.7}, 10^{10.5}]$, while the 
bottom panel presents those determined by satellites with $z=0$ stellar mass within the two more massive ranges considered, i.e. 
$M_{\star} [{\rm M}_{\odot}] \in [10^{10.5}, 10^{11.3}]$.

The sSFR evolution of low-mass satellites is clearly distinct from that of low-mass centrals (top panel of Fig.~\ref{fig:sSFRevolSatCen}); 
the decline of the sSFR towards lower redshifts is more pronounced for the former than for the latter.
This means that low-mass satellites become passive mainly because of the action of environmental processes,
reinforcing the conclusion inferred from the analysis of
Figs.~\ref{fig:fqz0_zinfall} and
~\ref{fig:fqzinfall_zinfall}.
The effect of TS on the DM subhaloes of satellites and the larger effects of RPS suffered by low-mass satellites (see Fig. 14 in Paper I) contribute to reduce the mass of their hot gas halo with the consequent reduction of the BH luminosity (top panel
of Fig.~\ref{fig:lumBHevolSatCen}), which is directly related to the efficiency of AGN feedback. 
The hot halo mass of centrals keeps growing towards lower redshifts as a result of cosmological gas accretion during halo-mass growth. Thus, according to equation~\ref{eq:BHgrowth-gascool}, the BH-growth rate through gas cooling also increases giving place to higher BH luminosity at lower redshifts. However, for low-mass centrals, these luminosities are not high enough to considerably reduce gas cooling ($L_{\rm BH}\sim 10^{48}\,{\rm erg}\,{\rm s}^{-1}$ at $z=0$). Consequently, this galaxy population remains mostly  
star-forming
at the present, as indicated by the solid line and associated shaded area in the top panel of Fig.~\ref{fig:sSFRevolSatCen}.

High-mass central galaxies are characterised by BH luminosities that are between $\sim 2$ (at $z=0$) and $\sim 3$ (at $z=1$) orders of magnitude larger than for low-mass ones (Fig.~\ref{fig:lumBHevolSatCen}). Thus,  AGN feedback becomes an efficient mass quenching process for more massive centrals,
as evident
from the abrupt decline of
the sSFR shown in the bottom panel
of Fig.~\ref{fig:sSFRevolSatCen}.
The stellar mass above which AGN feedback leads to SF quenching ($M_{\star} \gtrsim 10^{10.5}\,{\rm M}_{\odot}$) and the halo mass in which they reside ($M_{\rm halo}\gtrsim 10^{12}\, {\rm M}_{\odot}$)
are in agreement with the redshift-independent characteristic halo and stellar mass scales discussed by \citet{Henriques18}. 
The evolution of the sSFR and BH 
luminosity  
of high-mass central galaxies
are also closely followed
by high-mass satellites,
although with a lower normalization. 
Here, environmental processes contribute to the decline of star formation.
These results demonstrate that 
mass quenching keeps
playing a major role in the 
decline
of SF after high-mass galaxies become satellites.

\section{Satellite quenching times}
\label{sec:tq}

\begin{figure}
  \centering
  \includegraphics[width=\columnwidth]{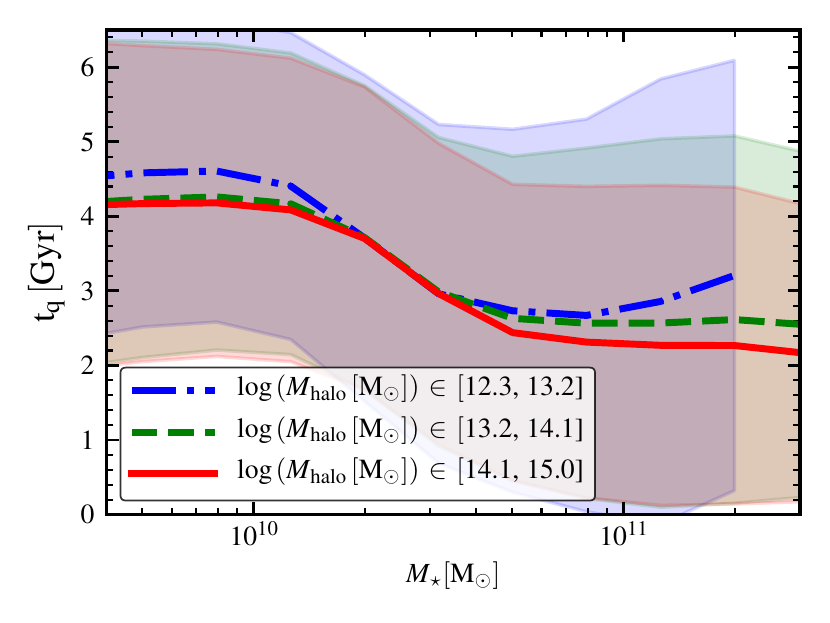}
  \caption{Mean values of quenching time-scale, $t_{\rm q}$, 
for satellite galaxies that are quenched at $z=0$ and 
were 
star-forming
at infall
as a function of their stellar mass. 
Different lines depict satellites residing within main host haloes of different 
present-day mass as indicated in the legend;
shaded areas represent the corresponding $1\,\sigma$ standard deviation around 
the mean. 
}
  \label{fig:tq-mstar-mhalo}
\end{figure}

\begin{figure*}
  \centering
  \includegraphics[width=2.0\columnwidth]{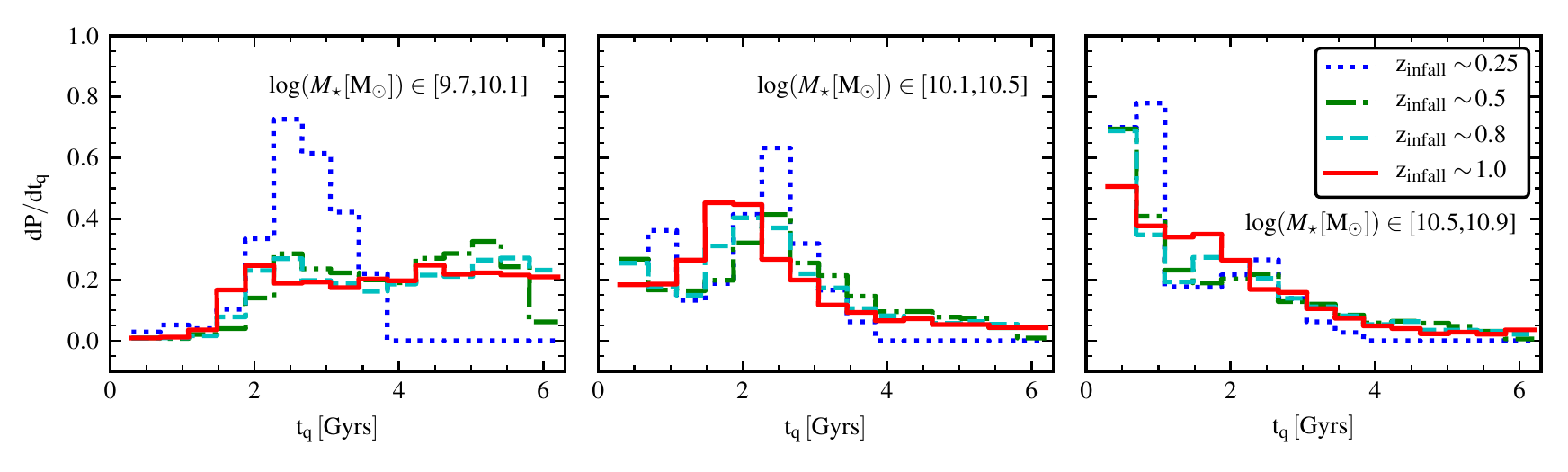}
  \caption{
Probability density function of quenching times, $t_{\rm q}$,
for the satellite population grouped in three stellar mass ranges
(different panels). Different style lines correspond to galaxies 
with different times of first infall which are comprised within a range of $\approx 1\,{\rm Gyr}$ around
four selected values of $z_{\rm infall}$ 
($z=0.25, 0.5, 0.8$ and $1$) as indicated in the legends,
which correspond to redshift ranges 
$z \in [0.2,0.3]$, $[0.45,0.6]$, $[0.7,0.9]$, $[0.95,1.2]$.
All galaxies within main host haloes of present-day mass
$M_{\rm halo} \geq 10^{12.3}\,{\rm M}_{\odot}$ 
are considered.
The most massive satellites are characterised by the lowest quenching
times ($t_{\rm q} \approx 0.05 - 2 \, {\rm Gyr}$).
The distribution of $t_{\rm q}$
peaks at $\approx 2 \, {\rm Gyr}$ for both satellites of intermediate mass and the lowest mass satellites that have been accreted
more recently ($z_{\rm infall}\in [0.2-0.3]$). 
Those 
low-mass
satellites that have been accreted earlier
are characterised by quenching times comprised within a broad range ($t_{\rm q} \approx 1.5 - 6 \, {\rm Gyr}$).
}
  \label{fig:tq}
\end{figure*}

Now we study the quenching times of those 
$z=0$
passive satellites that
are 
star-forming
at time of first infall, i.e. those that quench as satellites
by the contribution of both environmental and mass quenching. 
We estimate the quenching time, $t_{\rm q}$,
considering 
the period of time elapsed since 
$z_{\rm infall}$ and 
the moment in which the galaxy becomes passive,
identifying the snapshot
in the simulation in which 
the galaxy sSFR falls below the threshold 
$10^{-10.7}\,{\rm yr}^{-1}$.
This quenching time is defined 
in a similar way as in W13.
We present the relation between $t_{\rm q}$ and stellar mass
for satellites in main host haloes of different present-day mass in 
Fig.~\ref{fig:tq-mstar-mhalo}. 
The trends found are quite similar
to those shown by W13,
with 
low-mass
galaxies 
characterised by higher quenching times 
than 
more massive ones, 
and there being 
almost no secondary dependence on host halo mass.
Mean values of $t_{\rm q}$ predicted by our model are  
$\approx 4-5\,{\rm Gyr}$ for $M_{\star} \approx 10^{10} \,{\rm M}_{\odot}$ 
and $\approx 2-3\,{\rm Gyr}$ for $M_{\star} \approx 10^{11} \,{\rm M}_{\odot}$.
The scatter around these mean values is quite large, showing that 
low-mass
galaxies can achieve
quenching times as long as $6\,{\rm Gyr}$, consistent with W13.

The stellar mass dependence of the quenching time could be subject to bias if we consider 
the anticorrelation between the sSFR and stellar mass for galaxies 
in the star-forming main sequence.
By defining a quenching time as the time it takes to drop below an absolute value of sSFR, high-mass satellites would take less time to become passive because they have less distance to cover in sSFR space. 
However, the main sequence of star-forming satellites at different redshifts in our model
are rather flat (see fig. 7 in Paper I, thin dashed lines). Therefore,
the stellar mass dependence of the quenching time that we find is a robust result.
In order to confirm this, 
we have made the test of estimating the quenching time for galaxies defined as passive when they satisfy the condition
${\rm sSFR} < 10^{-12}\,{\rm yr}^{-1}$, that is, dropping the threshold adopted to define a galaxy as passive.
In this case, quenching times
for low-mass satellites become longer
by $\approx 0.8\,{\rm Gyr}$ but, contrary to expectations, quenching times for high-mass galaxies remain unaffected by the lower threshold, making the stellar-mass dependence of the quenching time even more pronounced. This 
is explained by the 
evolution of the sSFR for low- and high-mass satellites shown in 
Fig.~\ref{fig:sSFRevolSatCen}.
Low-mass satellites quench their SF in a more gradual way
than high-mass 
satellites. The latter are abruptly quenched reaching 
very low values of sSFR, 
so they are able 
to satisfy the more stringent condition imposed for being passive 
in the same short quenching time as 
required when the original threshold was applied. 

Regarding the halo mass independence, W13 argue that it reflects the fact that
$t_{\rm q}$ is measured 
since time of first infall, which involves preprocessing in smaller haloes;
the fraction of quiescent satellites that start quenching in another halo
varies from $\approx 0.15$ to $\approx 0.5$ depending on  current stellar mass and host halo mass
(see also \citealt{delucia2012}).
However, this lack of sensitivity of quenching times
to main host halo mass is also found by \citet{Oman16} even when 
they isolate environmental effects of the most recent host halo (within the range
$10^{13}-10^{15}\,{\rm M}_{\odot}$)
by measuring these time-scales since galaxies cross $2.5\,$\rvir~ 
of the current host (according to their definition of time of infall),
thus avoiding the time invested in pre-processing within other subhaloes. 
\citet{Oman16} find that quenching occurs after a delay time $\approx 3.5-5\,{\rm  Gyr}$ since infall
with quenching time-scales slightly shorter for higher mass galaxies.
Although the stellar mass dependence is weaker than the one found by W13,
time-scales obtained from these two works are of the same order of magnitude
once corrections for different definitions of infalling time are taken into account
\citep{Oman16}.
The general good agreement of \sagb~quenching times with values and
trends found by W13 and \citet{Oman16} is encouraging
and supports other related predictions of our model.

Fig.~\ref{fig:tq} shows the probability density functions
\footnote{Probability density function at the bin, normalized 
such that the integral over the range is equal to unity.}
of $t_{\rm q}$,
for satellites in 
three of the
stellar mass ranges analysed previously
(different panels), 
whose
redshifts of first infall are comprised within a range of 
$1\,{\rm Gyr}$ around
$z_{\rm infall}=0.25, 0.5, 0.8$ and $1$ (different line styles).
Since there is no halo mass dependence of $t_{\rm q}$,
all galaxies within main host haloes of present-day mass
$M_{\rm halo} \geq 10^{12.3}\,{\rm M}_{\odot}$
are considered.
The most massive satellites are characterised by the lowest quenching
times, with values of $t_{\rm q} \approx 0.05 - 2 \, {\rm Gyr}$.
Satellites of intermediate mass have a distribution of $t_{\rm q}$
that peaks at $\approx 2 \, {\rm Gyr}$.
This is also the case for the lowest mass satellites that have been accreted
more recently ($z_{\rm infall}\in [0.2-0.3]$). 
At first glance one could think that the latter result is biased by 
construction, 
since we are considering a set of galaxies that are 
star-forming
at infall 
and have quenched 
at $z=0$, so their time-scale for quenching is restricted. 
However, the fact that we do
find satellites satisfying all these criteria means that there is an efficient
mechanism acting at low redshifts that is able to quench SF of satellites.   
The quenching times of those low-mass satellites that have been accreted earlier cover a broad range from $\approx 1.5 \, {\rm Gyr}$ to $\approx 6 \, {\rm Gyr}$, with a slight preference for the longer times.

The combination of the previous analysis 
with the information on the relative proportion of galaxies accreted at different times allows us to explain the anti-correlation of the average values of $t_{\rm q}$ with stellar mass (Fig.~\ref{fig:tq-mstar-mhalo}).
For a given stellar mass range, the mean values of 
$t_{\rm q}$ 
are estimated considering satellites with all possible values of $z_{\rm infall}$.
The majority of 
$z=0$
passive satellites in the two highest 
stellar mass bins 
($M_{\star} > 10^{10.1} \,{\rm M}_{\odot}$)
have 
been accreted at low and intermediate redshifts 
($z_{\rm infall}\lesssim 0.5$),
whereas
the bulk of the current lowest mass galaxies considered here 
have become satellites much earlier.
Thus, early accreted low-mass
galaxies characterised by long quenching times outnumber the recently accreted ones, which reach shorter quenching times. Consequently,
the mean values of $t_{\rm q}$ become higher for less massive satellites.

\subsection{Discussion}
\label{sec:discussion}

Environmental effects such as~RPS and~TS
depend strongly on the orbit of the satellite
\citep[see e.g.][]{vollmer2001}. A galaxy on a more radial orbit, after passing
through the denser regions of the halo centre, will experience a strong~RP ($10^{-10}\,h^2\,{\rm dyn}\,{\rm cm}^{-2}$)
earlier than a galaxy which is initially in a slowly decaying circular orbit; it will also experience alternating periods of strong and weak~RP
($10^{-12}\,h^2\,{\rm dyn}\,{\rm cm}^{-2}$),
as its orbit
within the halo takes it from the dense central region to the lower-density
outskirts and back again. Another galaxy which is instead in a slowly decaying
circular orbit will experience an ever increasing~RP as it sinks towards denser
regions \citep{bdl2008, tecce10}. In the case of~TS, the mass loss also
depends strongly on orbital circularity
\citep[e.g.][]{tb2001,tb2004,zb2003,gan2010}, with most of the subhalo mass
being lost during pericentric passages.
The long quenching time-scales $t_{\rm q}$
of 
low-mass
satellites can be attributed to long dynamical friction time-scales
that keep 
low-mass
satellites in the outskirts of their main host haloes where RPS 
is less efficient.
Effectively, from the analysis of cosmological simulations,
\citet{Quilis17} show that massive satellites 
($M_{\star} > 10^{10}\,M_{\odot}$) are found at short halo-centric distances 
at low redshifts ($z\lesssim 0.5$), 
whereas
smaller systems are mainly located in the external
regions. This dichotomy in the stellar mass dependence of the average radial position is less pronounced at higher redshifts, with all galaxies being 
quite evenly distributed throughout the groups/clusters. 
It is worth noticing that low-mass satellites accreted at $z_{\rm infall} \gtrsim 0.5$ can have any quenching time within the
range $\approx 1 - 6$ Gyr (Fig.~\ref{fig:tq}). This result indicates that,  
depending on their conditions at infall,
many satellites are likely to have more eccentric orbits that bring them close 
to the halo centre characterised by higher values of RP and quench in short time-scales.

An important aspect to take into account on top of galaxy dynamics 
is the natural halo mass growth which is accompanied by an increase in the 
efficiency of RPS, 
giving place to faster quenching at lower redshifts.
This might account for the short quenching times 
($t_{\rm q}\approx 1-2\,{\rm Gyrs}$)
of high-mass satellites and recently accreted
low-mass
ones, as well as for the long quenching times of 
those satellites accreted at earlier times.

\subsection{Comparison with other works}

The values and the stellar mass dependence of the quenching times $t_{\rm q}$ obtained from our model are consistent with those estimated by W13. This trend with stellar mass
is also roughly recovered 
by \citet{Hirschmann14} who infer these time-scales from the requirement that 
their predicted quiescent fractions become consistent with observations,
since the semi-analytic model they use \citep{guo11} 
significantly overestimates the quiescent fraction of satellites and underestimate
that of centrals for the stellar mass and densities considered.

From the analysis performed on SDSS and 
3D-HST/CANDELS data, 
\citet{Fossati17} estimate quenching time-scales of passive satellites selected according to the their position in the 
rest-frame UVJ color-color diagram.  
Their definition of quenching time is similar to the one adopted in our work, i.e. the time elapsed since first infall till the satellite becomes passive. 
These times were obtained from mock 
catalogues constructed from a semi-analytic model \citep{henriques_2015} that match the number density and
redshift uncertainty of observed galaxies.
For quenched satellites at $z=0$ residing in low-mass haloes ($M_{\rm  halo} < 10^{13}\,{\rm M}_{\odot}$), they 
find quenching times of the order of $\approx 7 \,{\rm Gyr}$. The quenching times are lower ($\approx 5 \,{\rm Gyr}$) for satellites of more massive haloes, which the authors attribute to very massive haloes included 
in the SDSS dataset that exert larger environmental effects. 
Besides, for a given halo mass bin, these times have almost no dependence on stellar mass (see their fig.~21).
Our results, obtained for a population of 
$z=0$
quenched satellite galaxies, 
are not consistent with their findings at $z=0$
since the quenching times for our model galaxies show a much weaker dependence on halo mass, despite the fact that our halo-mass sample contains high mass clusters, and a clear trend with stellar mass.
The different approaches used to derive quenching time-scales make the identification of the source of discrepancies difficult.

On the other hand, 
the values of $t_{\rm q}$ obtained from our model and the stellar-mass and halo-mass trends they follow are consistent with the quenching times
obtained
for high redshift galaxies in the 3D-HST/CANDELS sample. For $z\sim 0.7-1.5$, \citet{Fossati17} find values of the order of $\approx 4-5 \,{\rm Gyr}$ for low mass galaxies 
and  $\lesssim 2\,{\rm Gyr}$ for
the most massive ones, with negligible dependence on halo mass within the considered range ($M_{\rm  halo}\lesssim 10^{14}\,{\rm M}_{\odot}$).
The latter values are larger than the
quenching time-scales found for clusters galaxies with $M_{\star} \gtrsim 10^{10.5}\,{\rm M}_{\odot}$ by \citet{Foltz18}, which are $\approx 1.5\, {\rm Gyr}$ and $\approx 1.24\,{\rm Gyr}$
at $z\sim 1$ and $z\sim 1.5$, respectively;
they classify galaxies as passive or 
star-forming
according to
dust-corrected rest-frame colours derived from spectral energy distribution 
 fitting assuming the
 delay-then-rapid quenching scenario (W13) to constraint the quenching time-scales.
A fair comparison with results from these works requires samples of model quenched satellites selected at high redshifts, which will be addressed in a near future.

\section{The delay-then-fade quenching scenario}
\label{sec:quenchscenario}

From the requirement of producing the correct sSFR distribution,
W13 propose the delay-then-rapid quenching scenario,
in which the times for the onset of SF quenching depend on the satellite mass
($t_{\rm q,delay} \approx 2 - 4\,{\rm Gyr}$) and, once started, 
the quenching takes place rapidly
with SFR declining exponentially 
with an e-folding time 
$\tau_{\rm Q,fade} \approx 0.8 \, {\rm Gyr}$ for $M_{\star} \approx 10^{9.7} \,{\rm M}_{\odot}$ and
$\tau_{\rm Q,fade} \approx 0.2 \, {\rm Gyr}$ for $M_{\star} \approx 10^{11.3} \,{\rm M}_{\odot}$.
In the delay phase of this two-stage model for satellite's SFR evolution, the SFR of an accreted star-forming galaxy is assumed to diminish gradually after infall in a similar way to the SFR of a central galaxy of the same stellar mass as the satellite at infall.
The quenching time-scales inferred by \citet{Oman16}  
also involve
a delay time until
the onset of quenching after infall ($\approx 3.5-5\,{\rm  Gyr}$) and a short time-scale for the fading of SF
($\lesssim 2\, {\rm Gyr}$). The latter was estimated by considering the time required by their galaxies to reach the fraction of passive satellites observed within galaxy clusters.

From Fig.~\ref{fig:sSFRevolSatCen}, we can see that the evolution of the sSFR of satellite galaxies in our model deviates  
from that of centrals some time after infall, and the decline of the sSFR of satellites becomes faster. This behaviour seems to be consistent with a two-stage model for the SFR evolution of satellites. However, it is not clear which physical processes are 
in action
in each of the two phases of SF quenching. 
The first stage in which the SFR of satellites evolves similarly to that of 
centrals (`delay phase') seems to be associated with the period of time required by the hot gas cooling to become inefficient after infall \citep[e.g.][]{Schawinski14}. 
This transition experienced by the gas cooling efficiency might define the beginning of 
the second stage in which the SFR of satellites declines more abruptly than that of centrals (`fading phase') because the cold gas reservoir
is no longer replenished by gas cooling from the hot halo. This phase might involve cold gas disc consumption through SF and/or removal through SN 
feedback and/or RPS, ending when the galaxy becomes passive.

\begin{figure}
  \centering
\includegraphics[width=\columnwidth]{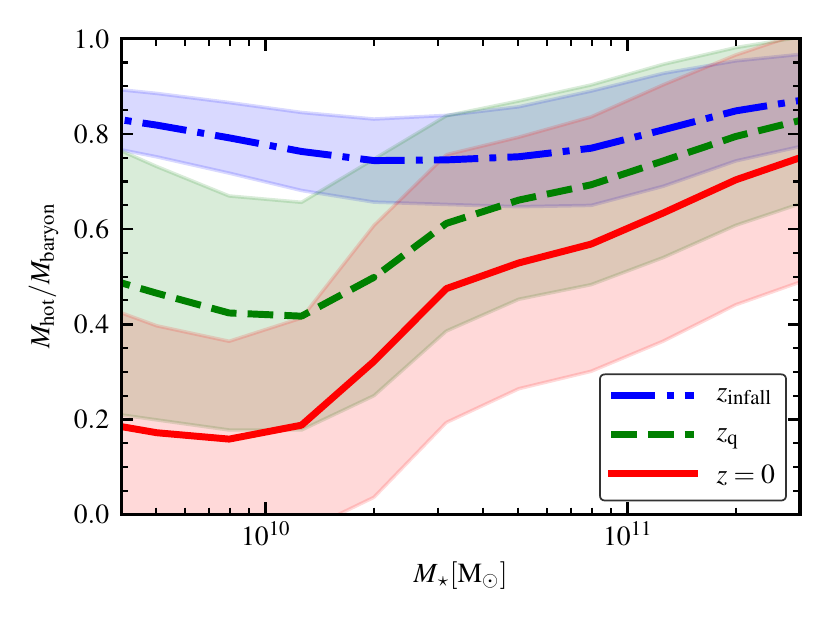}
  \caption{ 
Mean fraction of hot gas with respect to the total baryonic mass at different moments along the lifetime of $z=0$ passive satellites that are active at infall as a function of their $z=0$ stellar mass. These moments are: time of first infall ($z_{\rm infall}$; dashed dotted blue line), moment of SF quenching ($z_{\rm q}$; dashed green line), and the present epoch (z=0). The corresponding dashed areas denote the $1\sigma$ standard deviation. 
}
  \label{fig:fhot}
\end{figure}

In order to identify the physical mechanisms that play a relevant role in each of the two phases of the SFR decline, 
we analyse the mass content of the hot and cold gas reservoirs of $z=0$ passive satellites that are star-forming at infall considering different moments along their lifetime. 
Fig.~\ref{fig:fhot} shows the mean values of the fraction of hot gas with respect to the total baryonic mass, $f_{\rm hot}$,  
at infall (dashed-dotted line), at the moment of quenching, i.e, when their sSFR drops below the threshold adopted ($10^{-10.7}\,{\rm yr}^{-1}$; dashed line), and at $z=0$ (solid line),
as a function of their $z=0$ stellar mass. 
At time of first infall, $f_{\rm hot}\approx 0.8$ for any $z=0$ stellar mass; this fraction reflects the condition of central galaxies just prior to infall.
When high-mass satellites become passive, this fraction remains quite high ($f_{\rm hot} \approx 0.7$). For low-mass satellites, this fraction becomes lower ($f_{\rm hot} \approx 0.4$) but still consistent with the presence of a hot halo; the corresponding $1\sigma$ scatter indicates that  most of low-mass passive satellites have more than $20$ percent of hot gas at the moment of SF quenching.
We find that only $\approx 15$ and
$\approx 5$ percent of satellites with stellar mass $M_{\star} \approx 10^{10} \,{\rm M}_{\odot}$ and
$M_{\star} \approx 10^{11} \,{\rm M}_{\odot}$, respectively, quench their star formation after the hot gas has been reduced to less than $10$ percent of the total baryonic mass. Most of them have experienced at least one event of RPS of the cold gas before becoming passive.
These satellites 
are galaxies that have been accreted much earlier than the rest of the satellite population.
The mean values of their redshift of first infall 
vary from $z_{\rm infall} \approx 0.86$ to $\approx 0.94$, depending on the $z=0$ stellar mass range considered, while the corresponding values for those galaxies that quench their SF before hot gas depletion are comprised within the range $z_{\rm infall} \approx 0.52-0.78$. Therefore, the quenching times of this particular set of satellites
are longer ($t_{\rm q}\approx 5-6\,{\rm Gyr}$) than those typical of the bulk of the satellite population (see Figs.~\ref{fig:tq-mstar-mhalo} and~\ref{fig:tq}). 

\begin{figure}
  \centering
\includegraphics[width=\columnwidth]{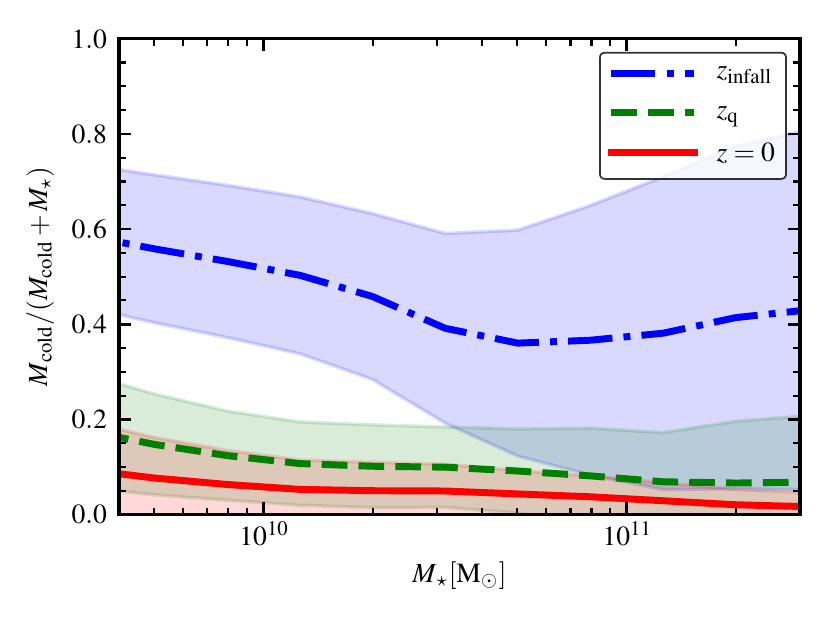}
  \caption{ 
Mean fraction of cold gas with respect to the sum of cold gas and stellar mass at different moments along the lifetime of $z=0$ passive satellites that are active at infall as a function of their $z=0$ stellar mass. These moments are: time of first infall ($z_{\rm infall}$; dashed dotted blue line), moment of SF quenching ($z_{\rm q}$; dashed green line), and the present epoch (z=0). The corresponding dashed areas denote the $1\sigma$ standard deviation.
}
  \label{fig:fcold}
\end{figure}

One important implication of the aforementioned results is that
the complete removal of the hot gas halo is not a requirement neither for the onset of SF quenching (beginning of a faster decline of the SFR) nor for the SF quenching itself.   
Gas cooling, which is linked to the properties of the hot gas halo, namely, mass, metallicity and density profile \citep[e.g.][]{springel2001},
may become inefficient even when a substantial amount of hot gas is still available. 
The hot gas halo kept by galaxies after infall is mildly reduced by RPS for high-mass satellites (see fig.~14 in Paper I), but gas cooling is considerably suppressed by the action of AGN feedback (equation~\ref{eq:cool}). This reduction of the gas cooling efficiency for more massive star-forming galaxies becomes evident in 
parsec-scale hydrodynamical simulations \citep{Armillotta16}.
This is consistent with the dominance of mass quenching over environmental quenching in the SFR decline of high-mass satellites (see Figs.~\ref{fig:sSFRevolSatCen}~and \ref{fig:lumBHevolSatCen}).
In low-mass satellites, 
gas cooling 
may become inefficient once the hot gas reservoir has been sufficiently reduced, although not necessarily depleted, by the stronger effect of RPS and gas cooling itself. 
Further reduction of the hot gas halo after SF quenching is the result of the action of RPS that contributes to determine the fractions of hot gas achieved at $z=0$ ($f_{\rm hot} \approx 0.2$ for $M_{\star} \approx 10^{10}\,M_{\odot}$ and $f_{\rm hot} \approx 0.6$ for $M_{\star} \approx 10^{11}\,M_{\odot}$, Fig.\ref{fig:fhot}). Clearly, this effect dominates over any possible replenishment with gas reheated by SN feedback. 
Note that satellites do not accrete hot gas cosmologically in our model, a common feature in SAMs \citep[e.g.][]{henriques_2015,Stevens17}. 

Fig.~\ref{fig:fhot} shows that the bulk of the $z=0$ passive satellite population experiences SF quenching that is not associated to the action of RPS of the cold gas disc,
which in our model takes place only when this component is not longer protected by the hot gas halo.
This result is supported by qualitatively similar conclusions obtained from 
cosmological chemodynamical simulations \citep{Kawata08} and  the
semi-analytic model \textsc{dark sage}  \citep{Stevens17}. This SAM 
also includes the effect of RPS on both the cold-gas disc and the hot-gas halo, with the former being shielded by the presence of the latter, although the corresponding implementations vary with respect to those in our model.

Fig.~\ref{fig:fcold} shows 
the mean values of the fraction of cold gas with respect to the sum of cold gas and stellar mass, $f_{\rm cold}=M_{\rm cold}/(M_{\rm cold}+M_{\star})$, for the same satellite population and stages as in Fig.~\ref{fig:fhot}. Mean values at time of first infall vary from $f_{\rm cold}\approx 0.6$ for $M_{\star} \approx 10^{10}\,M_{\odot}$ to $f_{\rm cold} \approx 0.4$ for $M_{\star} \approx 10^{11}\,M_{\odot}$; the 1-$\sigma$ dispersion is larger for more massive satellites, which indicates that some high-mass galaxies are already gas-poor when they are accreted, with levels of SF that are likely close to the limit of quenching. 
This is a clear sign of the action of mass-quenching processes on high-mass galaxies prior to infall (see Fig.~\ref{fig:fqzinfall_zinfall}).
This also explains the lower values of the quenching times, $t_{\rm q}$, of more massive satellites (see Figs.~\ref{fig:tq-mstar-mhalo}~and \ref{fig:tq}). All satellites, regardless of their $z=0$ stellar mass, have low cold gas fractions by the time they become passive ($f_{\rm cold} \approx 0.1-0.2$); their cold gas  
has been gradually consumed by star formation and/or removed through outflows produced by SN feedback. 
The fraction $f_{\rm cold}$ 
continues decreasing beyond
the quenching time $t_{\rm q}$
up to the present epoch by the remaining modest star formation taking place
and/or the action of 
RPS on the cold gas disc if it is no 
longer shielded by the hot gas halo. 

Since gas cooling efficiency plays a key  role in the SFR decline, we estimate the rates of gas cooling at specific moments, following the analysis applied to the hot and cold gas fractions (see Figs.~\ref{fig:fhot} and~\ref{fig:fcold}).  
Satellites of any $z=0$ stellar mass have, on average, gas cooling rates $\lesssim 5\,{\rm M}_{\odot}\,{\rm yr}^{-1}$ when becoming passive.
This result means that there is not a sharp cut-off in the cold gas supply.
Instead, the gas cooling rate becomes progressively lower departing from values within the range $\approx 30-40\,{\rm M}_{\odot}\,{\rm yr}^{-1}$ when star-forming galaxies are accreted. 
From the estimation of the evolution of median gas cooling rate for the same satellite population considered in Fig.~\ref{fig:sSFRevolSatCen}, we can see (not shown here) 
a transition to a faster reduction of the gas cooling rate at $\approx 20\,{\rm M}_{\odot}\,{\rm yr}^{-1}$, which is connected with the beginning of the fading phase where the SFR decline is faster. 
Thus, the results of our model indicate that the fading phase begins when the gas cooling rate has been reduced by $\approx 50$ percent with respect to its value at infall. 

Our results support a two-stage quenching scenario for satellite galaxies, where the length of time of the delay and fading phases are comprised within the time $t_{\rm q}$, which characterises the whole quenching process. 
We estimate the length of time of the fading phase, $t_{\rm q,fade}$, as  
the period of time 
elapsed since the gas cooling rate is reduced to half the value it has at infall
until the satellite becomes passive. 
We find that $t_{\rm q,fade}\approx 1.5\,{\rm Gyr}$ for satellites of any $z=0$ stellar mass, with a very mild trend to be shorter for more massive galaxies.
This result is representative of a large percentage of the satellite galaxies in our sample. The cooling-rate-based condition of transition from the delay phase to the fading one is satisfied by 
$\approx 90$ percent of satellites with stellar mass $M_{\star} \approx 10^{10.5} \,{\rm M}_{\odot}$. This percentage decreases for satellites with both lower and higher stellar masses, and reaches values as low as $\approx 60$ and $\approx 40$ percent, respectively. Those satellites that do not fulfil this requirement are galaxies that already have very low cooling rates when they are accreted ($\approx 5\,{\rm M}_{\odot}\,{\rm yr}^{-1}$).

For those satellites that go through  both the delay and fading phase, we estimate the length of time of the delay phase, $t_{\rm q,delay}$, from the difference between the mean values of
$t_{\rm q}$ (Fig.~\ref{fig:tq-mstar-mhalo}) and $t_{\rm q,fade}$ for each stellar mass range.
The delay time varies from $t_{\rm q,delay}\approx 3\,{\rm Gyr}$ for low-mass satellites to $t_{\rm q,delay}\approx 1\,{\rm Gyr}$ for high-mass ones. 
Taking into account the scatter in the estimation of these times, 
we find that the values and stellar mass dependence of $t_{\rm q,delay}$ are in agreement with those obtained by W13. Mean values of $t_{\rm q,fade}$ are also consistent with the e-folding time over which the SFR
fades, $\tau_{\rm Q,fade}$, estimated by W13,
although our model does not predict statistical significant lower values for more massive satellites.
We can  conclude that the SF quenching of $z=0$ passive satellites can be described by a two-stage quenching scenario, characterised by a delay and a fading phase. The fading time is largely independent of $z=0$ stellar mass. It  is shorter than the delay time for low-mass galaxies, 
whereas the opposite situation occurs for high-mass ones. Therefore, we consider that the term `delay-then-rapid' proposed by W13 to dub the SF quenching scenario is not representative of the SF history of $z=0$ passive satellites with high stellar mass. Thus, we prefer to use the more 
inclusive
term {\em delay-then-fade} to fairly describe all the possible situations.

\subsection{Discussion}

Based on the physical processes implemented in the \sag~model (gas cooling, star formation, SN and AGN feedback, RPS of the hot and cold gas phases),
we have identified the relative impact of mass and environmental quenching on the SF history of $z=0$ passive satellites. We find that the SF quenching of satellites is well described by a delay-then-fade quenching scenario. 
The rate of gas cooling from the hot halo plays a decisive role in the beginning of the fading phase. It is mainly determined by an internal process (AGN feedback) in high-mass satellites, and by an environmental process (RPS of the hot gas halo) in low-mass ones.
The cut-off of the cold gas replenishment
by gas cooling, 
regardless of which physical
process may actually be responsible for it,
is named `strangulation'
\citep[e.g.][]{Peng15}.
It is important to introduce a note of caution regarding the meaning attributed to this term in different works in the literature. 
The works that have introduced this concept \citep{larson80,balogh2000,balogh00b} consider that strangulation also involves the stripping of the hot gas halo of  satellite galaxies, a convention followed in subsequent works \citep[e.g.][]{Kawata08}. The evidences that the hot gas removal takes place in a few Gyrs after accretion \citep{Rasmussen06,Vijayaraghavan15} justify the gradual starvation scenario implemented in \sag~(see Sec.~\ref{sec:model}). However, in previous versions of our model, as well as of other SAMs \citep{kauffmann93,weinmann2006b, croton2006}, the modelling of the strangulation mechanism has been oversimplified by assuming that the hot gas halo is removed instantly when a galaxy becomes a satellite. Thus, in the context of SAMs, the term `strangulation' implies this crude formulation of the process.

\citet{Peng15} propose that strangulation is the primary mechanism for shutting down star formation.
They 
find that the
mass-dependent metallicity difference between quiescent 
and star-forming galaxies in SDSS
can be very well reproduced by a close-box model that assumes that
galaxies become passive after the cold gas supply is halted.
However, the results from our model suggest that strangulation, 
in the general sense used by \citet{Peng15}, 
is a very strict condition to define the beginning of the SF fading because it is not necessary a complete suppression of gas cooling but only a more pronounced reduction of the cooling rates.
Besides, even assuming strangulation, they obtain 
a typical time-scale for SF quenching of $\approx 4\,{\rm Gyr}$ (largely independent of stellar mass), 
which is longer by $2.5\,{\rm Gyr}$ than our predictions for $t_{\rm q,fade}$.
This disagreement points to the fact that a simple close-box model 
is not adequate to capture all the complex physical processes, both internal
and environmental, that affect the star formation activity of a galaxy.
Although this objective is better accomplished by \sag, it is worth noticing that our model is not able to recover the observed metallicity difference between star forming and quiescent galaxies.
We find an increasing trend of stellar metallicity with stellar
mass in both cases, but not a significant difference between
the metallicities of these two populations at a given stellar mass.
This could be attributed to the fact that we discriminate galaxies in passive
and 
star-forming
according to their sSFR instead of using their colours, as
in \citet{Peng15}. Therefore, many satellites classified as 
star-forming
might be close
to be quenched and their metallicities would be 
pretty similar to those of passive
galaxies. This issue deserves a deeper analysis.

The results of our model are in  
line with 
the scenario of `overconsumption' proposed
by \citet{McGee14}, based on the analysis of satellite quenching
times at a range of redshifts derived from observations. In this scenario, SF quenching of satellite galaxies is driven by secular outflows
once cosmological accretion of gas is halted, and  
satellites become passive long before orbit-based gas stripping, such as RPS of the hot halo or of the cold gas disc, can have some impact on the quenching process.
The predictions of \sagb~agree partially with the later aspect.
Although our model supports the fact that RPS of the cold gas plays 
a secondary role or even has negligible
effect on the SF quenching, the RPS
on the hot gas halo 
does play an important role since 
this process contributes to reduce the hot gas reservoir of low-mass satellites which drives the decline of the gas cooling rate.

In the context of their overconsumption model, \citet{McGee14} find that 
a constant mass-loading of the
wind can reproduce the evolution of the quenching time-scales of the delay phase as compiled from
the literature (including the local value given by W13, 
which is of the same order of magnitude as the one inferred from \sagb).
\citet{Oman16} show that 
this simple model is in conflict with
the stellar mass dependence of the quenching time-scale. This conclusion is supported by our model, 
which is able to reproduce such a dependence based on a  
new feedback scheme
that involves a
redshift dependent mass-loading factor 
(see eq.~\ref{eq:feedfire}),
a feature that has demonstrated to be crucial in reproducing several
observed galaxy properties (Paper I; \citealt{Collacchioni18}).

\section{Summary and conclusions}
\label{sec:conclu}

We have analysed a galaxy catalogue generated by applying
the updated version of our semi-analytic model of galaxy
formation
\sag, described in detail in Paper I, on the cosmological
\textsc{MultiDark} $1\,h^{-1}\,{\rm Gpc}$ MDPL2 simulation with the aim
of contributing to our
understanding of the relative role of environmental and mass quenching
processes on satellite galaxies of different present-day stellar mass
hosted by DM haloes of different mass.  We also estimate the
quenching time-scales involved.
The latest improvements implemented in \sag~include
a robust model of environmental effects through the action of RPS and TS
coupled to the integration of the orbits of orphan satellites,
and a higher efficiency of SN feedback allowed by an explicit
redshift dependence of the reheated and ejected mass.
In Paper I, we have demonstrated that a variant of the model referred to as
\sagb~allows us to
achieve good agreement with observational results for
several galaxy properties at both low and high redshifts. 
In particular, the agreement of the predictions of \sagb~with the observed fraction of currently passive satellites as a function of stellar mass, halo mass and halo-centric distance makes this model suitable to carry out the present work.

We analyse subsamples of galaxies
selected according to their stellar mass, main host halo mass and
time of first infall. Our main conclusions can be summarised as follows:

\begin{itemize}

\item 
From the analysis of the relative importance of mass quenching and environmental quenching for local quenched satellites of different stellar mass, we find
$M_{\star} \approx 10^{10.5} \,{\rm M}_{\odot}$ to be the mass scale where mass quenching becomes important. This is also a characteristic mass scale for quenching in central galaxies \citep{Henriques18}.
Environmental processes, 
on the other hand, dominate the SF quenching of 
low-mass
satellite galaxies
($M_{\star} \lesssim 10^{10.1} \,{\rm M}_{\odot}$).
This picture is 
consistent with 
the results of 
previous works 
(\citealt{vandenBosch08}, \citealt{Peng10}, W13, \citealt{Lin14}, \citealt{Kawinwanichakij17}, \citealt{CochraneBest18}).
These conclusions  
are inferred
from the following results:

\begin{itemize}

\item
Galaxies of any stellar mass that have been satellites for more than $\approx 8\, {\rm Gyr}$ ($z_{\rm infall} \gtrsim 1$) are characterised by 
similarly 
high
values of 
$z=0$
quenched fractions
($fq_{\rm z0} \approx 0.8 - 0.95$).
Such high quenched fractions suggest that the time elapsed since first infall is enough for 
the combined action of
mass and environmental processes to fully quench 
satellite
galaxies.

\item 
Low-mass
satellites
($M_{\star} \lesssim 10^{10.1} \,{\rm M}_{\odot}$)
have not suffered mass quenching while being centrals
as evidenced by the null values of the
corresponding
quenched fraction at time of first infall ($fq_{\rm infall}$; 
Fig.~\ref{fig:fqzinfall_zinfall}).
Their 
low mass prevents them from being quenched by self-regulating processes
such as AGN feedback or disc instabilities.
The same is true after infall, i.e. the quenched fractions at $z=0$ result
solely from environmental processes.
Values of $fq_{\rm z0}$ are larger for galaxies that
have been accreted earlier (higher $z_{\rm infall}$)
because they have been satellites for longer periods of time, thus being affected by environmental quenching mechanisms for longer
(Fig.~\ref{fig:fqz0_zinfall}).

\item  High-mass satellites ($M_{\star} \gtrsim 10^{10.5} \,{\rm M}_{\odot}$) are more
likely to be quenched 
prior to infall.
At a given redshift of first infall,
the fractions of quenched satellites $fq_{\rm infall}$ are higher for more massive galaxies (Fig.~\ref{fig:fqzinfall_zinfall}).
At early accretion epochs ($z_{\rm infall}\approx 1.5$),
$\approx 30$ percent of galaxies with local stellar mass
$M_{\star} \gtrsim 10^{10.9} \,{\rm M}_{\odot}$ 
accreted by high-mass 
haloes ($M_{\rm vir}[{\rm M}_{\odot}] \in [10^{14.1}, 10^{15.}]$) are quenched. 
The fractions $fq_{\rm infall}$ increase for lower values of $z_{\rm infall}$,
which may result from the combination of the stellar mass growth of a galaxy prior to infall
and the time elapsed under the action of  mass quenching processes while being central.

\item Mass quenching plays a major role in the 
decline of the SF for high-mass galaxies after infall.
Within the same stellar and host halo mass 
range 
defined at a given time of first infall,
satellites and centrals follow a similar evolution of the sSFR and BH luminosity,  directly related with the efficiency of AGN feedback
(Figs.~\ref{fig:sSFRevolSatCen} and~\ref{fig:lumBHevolSatCen}). 
These quantities are slightly smaller for satellites as a consequence of the additional effect of 
RPS, which dominates among the
environmental processes included
in our model and exerts milder effects on more massive galaxies (Paper I).

\end{itemize}

\item
For a given 
$z=0$
stellar mass, SF quenching mechanisms are less efficient in galaxies 
accreted by lower mass haloes 
both prior to and after their first infall.
The stellar mass at infall is smaller for satellites of less massive haloes
as a consequence of the different stellar-mass growth rates that characterise galaxies of the same $z=0$ stellar mass in different environments. 
Therefore, at a given time, galaxies accreted by lower mass haloes experience milder mass quenching 
prior to first infall (lower values of $fq_{\rm infall}$) than a population with the same $z=0$ stellar mass infalling in more massive haloes (Fig.~\ref{fig:fqzinfall_zinfall}).
Moreover, mass quenching after infall is also reduced, which combined with 
the milder environmental effects exerted by lower mass haloes produces lower values of $fq_{\rm z0}$.
Both fractions $fq_{\rm z0}$ and $fq_{\rm infall}$ are reduced for less massive haloes at any stellar mass. However,
$fq_{\rm z0}$ is more strongly reduced, and thus the ratio 
$fq_{\rm infall}/fq_{\rm z0}$, which gives the fraction of 
galaxies quenched at $z=0$
that are already quenched at first infall, takes 
larger values
for galaxies residing in host haloes of lower mass 
(Fig.~\ref{fig:fq-at-infall-z0-ms-mh}). 
The interpretation of this trend may lead to a conclusion opposite to the one drawn from our analysis, i.e. 
that quenching prior to infall is more important in less massive
haloes, as the one discussed in W13.
\\

\item The quenching times of 
$z=0$
passive satellites that were 
star-forming
at first infall are anti-correlated with their present-day stellar mass (Fig.~\ref{fig:tq-mstar-mhalo}), consistent with results from W13:
the average values of $t_{\rm q}$ are  
$\approx 4-5\,{\rm Gyr}$ for $M_{\star} \approx 10^{10} \,{\rm M}_{\odot}$ 
and $\approx 2-3\,{\rm Gyr}$ for $M_{\star} \approx 10^{11} \,{\rm M}_{\odot}$.
These average values of $t_{\rm q}$ are estimated including satellites with all possible values of $z_{\rm infall}$.
The anti-correlation with $z=0$ stellar mass arises because 
early accreted 
low-mass 
galaxies 
achieve quenching times as long as $\approx 6\,{\rm Gyr}$ (Fig.~\ref{fig:tq}), and these galaxies
outnumber the recently accreted ones, characterised by shorter quenching times.
\\

\item 
Overall, we can characterise the SF quenching process of 
$z=0$ passive satellites as consisting of two stages. During the first one,  which is measured from 
the moment that a 
star-forming
galaxy becomes a satellite,
the gradual decline of the SFR resembles that of centrals of the same stellar mass as the satellites at infall. High levels of SF are sustained by high rates of gas cooling that only experience a mild reduction during this stage; their values at infall are 
$\approx 30-40\,{\rm M}_{\odot}\,{\rm yr}^{-1}$.
When the cooling rate reaches half its value at infall ($\approx 20\,{\rm M}_{\odot}\,{\rm yr}^{-1}$, on average),
there is a transition from a slow to a rapid decline of the gas cooling rate, as suggested by the evolution on the sSFR in  Fig.~\ref{fig:sSFRevolSatCen},
which denotes the end of the delay phase.
This process takes place in a delay time  
that ranges from  
$t_{\rm q,delay} \approx 3\,{\rm Gyr}$ for low-mass
satellites ($M_{\star} \lesssim 10^{10}\,{\rm M}_{\odot}$)
to $\approx 1\,{\rm Gyr}$ for 
high-mass ones ($M_{\star} \approx 10^{11}\,{\rm M}_{\odot}$).
In the second stage,
the SFR declines faster until the satellite becomes passive (see the behaviour of the sSFR in Fig.~\ref{fig:sSFRevolSatCen}).
SF fades because the cold gas 
supply is reduced at a faster rate (reaching values as low as 
$\approx 5\,{\rm M}_{\odot}\,{\rm yr}^{-1}$ by the time the satellites are quenched) rather than being halted at the beginning of this fading phase as assumed in the strangulation quenching scenario \citep{Peng15}.
The cold gas disc is gradually consumed mainly through SF and/or removal through SN feedback. This process occurs in a fading time 
$t_{\rm q,fade} \approx 1.5\,{\rm Gyr}$, regardless of stellar mass. 
Our model is only consistent with the delay-then-rapid quenching scenario proposed by W13 for low-mass satellites. Since the delay time is shorter for more massive satellites and the fading time is largely independent of stellar mass, we find that the SF history of all $z=0$ passive satellites is better described by a {\em delay-then-fade} quenching scenario.
\\

\item Environmental processes have an important role during the delay for 
the onset of SF quenching for low-mass satellites.
The gradual removal of the hot gas halo through RPS
is directly influenced by the 
orbital evolution of galaxies and the mass growth of DM haloes.
The relevance of this process diminishes for high-mass satellites
which keep a large fraction of the hot gas reservoir by the time they become passive ($f_{\rm hot}\gtrsim 0.6$ ). These fractions are smaller for low-mass galaxies (mean values of $f_{\rm hot} \approx 0.4$; fractions can be as low as $\approx 0.2$ considering the scatter) but still consistent with the presence of a hot halo (Fig.~\ref{fig:fhot}). The larger reduction of the hot gas mass in less massive satellites yields 
to the decrease of their gas cooling rates. For high-mass satellites, this decrease occurs as a consequence of AGN feedback.  
RPS of the cold gas does not play any role in the fading of SF. It only acts on those satellites that have lost their protective hot gas halo, and contributes towards reducing even more the cold gas fraction after quenching (Fig.~\ref{fig:fcold}).

\end{itemize}

It is worth noting that either mass quenching or starbursts triggered by mergers could be responsible for
the SF quenching of galaxies while being centrals. The relative importance of these processes will be examined in 
another work. Furthermore, we plan to extend the analysis presented here to high redshifts. The dependence of
the quenching time-scales on redshift has been discussed recently in several observational studies \citep{Balogh16,Fossati17,Foltz18}. Comparison with
results provided by them will allow both to test our model and to help to our understanding of the SF quenching over cosmic time.

\section*{Acknowledgements}
We thank the referee for the constructive report that improved the quality of the manuscript.
The authors gratefully acknowledge the Gauss Centre for Supercomputing e.V. 
(www.gauss-centre.eu) and the Partnership for Advanced Supercomputing in Europe 
(PRACE, www.prace-ri.eu) for funding the \textsc{MultiDark} simulation project 
by providing computing time on the GCS Supercomputer SuperMUC at Leibniz 
Supercomputing Centre (LRZ, www.lrz.de). The MDPL2 simulation has been performed 
under grant pr87yi. 
This work was done in part using the Geryon computer at the
Center for Astro-Engineering UC, part of the BASAL PFB-06, which received
additional funding from QUIMAL 130008 and Fondequip AIC-57 for upgrades.
SAC acknowledges funding from {\it Consejo Nacional de Investigaciones
Cient\'{\i}ficas y T\'ecnicas} (CONICET, PIP-0387), {\it Agencia Nacional
de Promoci\'on Cient\'ifica y Tecnol\'ogica} (ANPCyT, PICT-2013-0317), and {\it Universidad Nacional de La Plata} (G11-124),
Argentina.
TH and CVM acknowledge CONICET, Argentina, for their supporting fellowships.
AO acknowledges support from project AYA2015-66211-C2-2  of  the Spanish Ministerio de Econom\'ia, Industria y Competitividad.
This project has received funding from the European Union's Horizon 2020 Research and Innovation Programme under the Marie Sklodowska-Curie grant agreement No 734374.

%%%%%%%%%%%%%%%%%%%%%%%%%%%%%%%%%%%%%%%%%%%%%%%%%%



\bibliographystyle{mnras}
\bibliography{references}

%%%%%%%%%%%%%%%%%%%%%%%%%%%%%%%%%%%%%%%%%%%%%%%%%%

%%%%%%%%%%%%%%%%% APPENDICES %%%%%%%%%%%%%%%%%%%%%

%\appendix



%%%%%%%%%%%%%%%%%%%%%%%%%%%%%%%%%%%%%%%%%%%%%%%%%%

% Don't change these lines
\bsp	% typesetting comment
\label{lastpage}
\end{document}